\documentclass[%
 aip,
 amsmath,amssymb,
 reprint,%
]{revtex4-1}

\usepackage{graphicx}
\usepackage{dcolumn}
\usepackage{bm}

\usepackage[utf8]{inputenc}
\usepackage[T1]{fontenc}
\usepackage{mathptmx}
\usepackage{newtxtext}
\usepackage{newtxmath}
\usepackage{etoolbox}
\usepackage{bm} 

\makeatletter
\def\@email#1#2{%
 \endgroup
 \patchcmd{\titleblock@produce}
  {\frontmatter@RRAPformat}
  {\frontmatter@RRAPformat{\produce@RRAP{*#1\href{mailto:#2}{#2}}}\frontmatter@RRAPformat}
  {}{}
}%
\makeatother
\begin{document}

\preprint{AIP/123-QED}

\title{Rayleigh-B\'enard-Marangoni convection in binary fluids - Effect of miscibility on oscillatory mode}
\author{Anubhav Dubey}
 \email{anubhav.dubey@u-bordeaux.fr}
\affiliation{ 
Univ. Bordeaux, CNRS, Bordeaux INP, I2M, UMR 5295, F-33400, Talence, France
}%
\author{Saumyakanta Mishra}%
\affiliation{ 
Engineering Mechanics Unit, Jawaharlal Nehru Center for Advanced Scientific Research, Jakkur, Bangalore, 560064, Karnataka, India 
}%

\author{S. V. Diwakar }
\affiliation{
Engineering Mechanics Unit, Jawaharlal Nehru Center for Advanced Scientific Research, Jakkur, Bangalore, 560064, Karnataka, India 
}%

\author{Sakir Amiroudine}%
\affiliation{ 
Univ. Bordeaux, CNRS, Bordeaux INP, I2M, UMR 5295, F-33400, Talence, France
}%

\date{\today}

\begin{abstract}
The pair of fluids, FC72- 1cSt silicone oil, exhibiting temperature-sensitive miscibility gap is considered in this study to investigate the classical Rayleigh-B\'enard-Marangoni (RBM) instability. The system of fluids is considered at distinct temperatures to elucidate the effect of the degree of miscibility between the two fluids. We employ a modified phase-field model to track the evolution of the RBM instability as the temperature of the system is varied. The proposed model correctly initializes the concentration profile and also properly leads to a miscibility-dependent surface tension. A spectral-collocation-based method is employed to solve the linearized governing equations that help investigate the onset characteristics of the convection, i.e., both the critical values and the mode of convection onset, which could be either oscillatory or stationary. The results reveal the dependence of the window of oscillatory convection on the degree of miscibility between the two fluids. The coupling between the Korteweg stresses and the degree of miscibility governs the thermal energy required to provoke the system into the convective state. The onset of RBM flow is analyzed parametrically, with the successive rise in the strength of the Marangoni component.      
\end{abstract}

\maketitle

\section{\label{sec:level1}Introduction}

Binary fluids with temperature-sensitive miscibility gap have received renewed scientific attention owing to the discovery of several potential applications such as targeted drug delivery \cite{Schmaljohann2006}, extraction and separation of bio-active compounds \cite{Kohno2011, Ventura2017}, enhancement of liquid-liquid mass transfer \cite{Dessimoz2008}, liquid-crystal micro-droplet formation \cite{Patel2021} to enumerate a few. For such a fluid pair, the relative degree of miscibility between the two fluids is a function of temperature. The directional evolution of the miscibility gap allows binary fluids to be categorised as either fluids with lower-critical solution temperature (LCST) or upper-critical solution temperature (UCST). The UCST is the maximum temperature until the two fluids have a discernible interface separating the fluid-1 dominated region from the fluid-2 dominated region. The concentration gradient in the interfacial region thus varies as a function of the temperature. Consequently, the Korteweg \cite{Joseph1990} stresses, arising due to the concentration gradient, can be manipulated by varying the temperature of the system. This opens up the avenue for microfluidic flow pattern tuning \cite{Fornerod2020} of binary fluids with explicit control of temperature. Further, the sensitivity of capillary effects induces thermo-capillary flow, which governs the microdroplet formation while the system undergoes phase separation \cite{Aibara2020}. Thus, a thorough understanding of the spatio-temporal evolution of binary fluids is necessary. In light of the same, the current study explores the classical Rayleigh-B\'enard-Marangoni (RBM) instability at distinct system temperatures to reveal the onset criteria of convection, using a specific system involving FC-72 and silicone oil. Before venturing into further details, a brief overview of the rich double-layer RBM instability literature is provided below.    

\begin{figure}
\centering 
\includegraphics[width=0.45\textwidth]{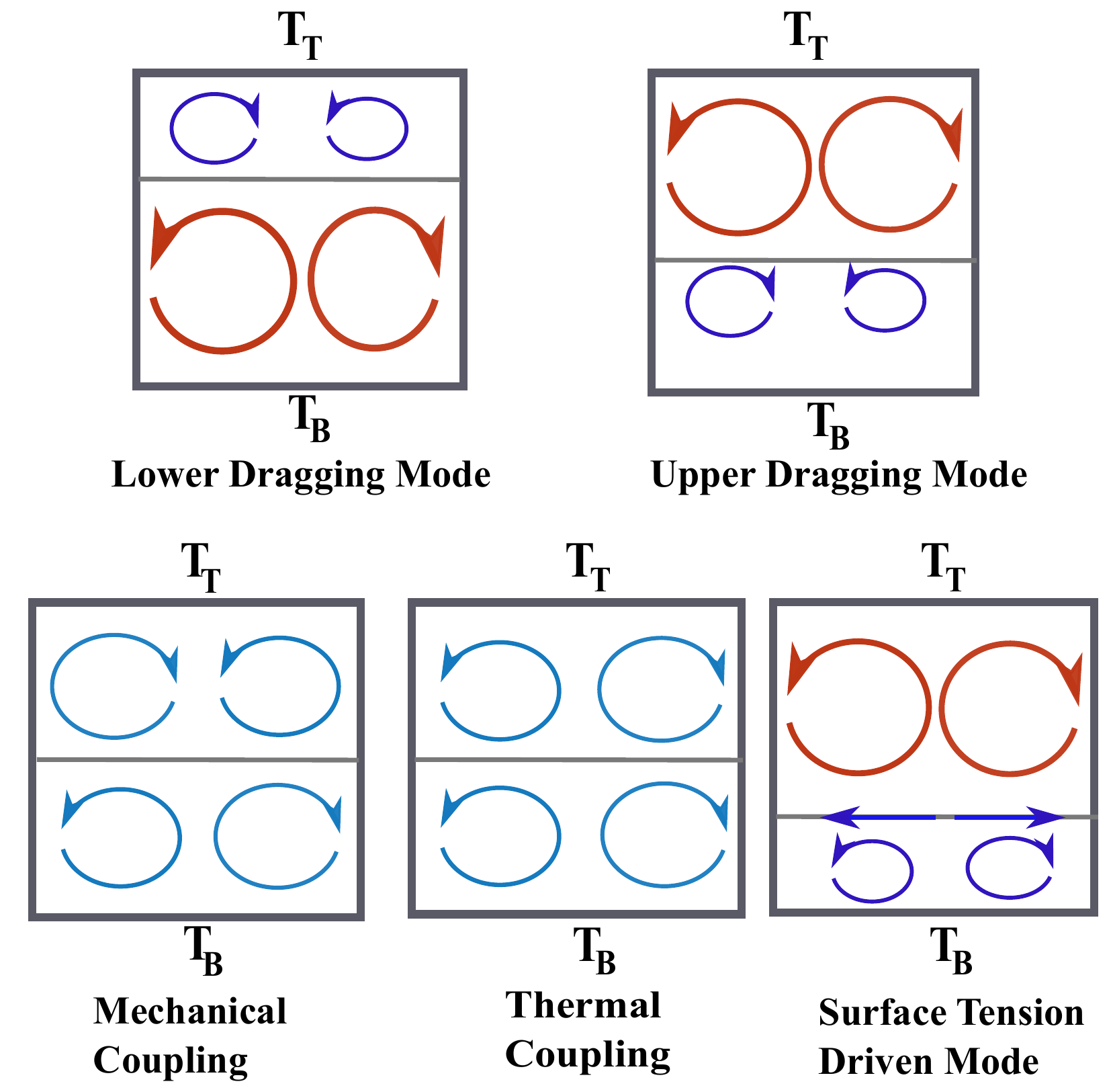} 
\caption{Qualitative sketch of non-oscillatory modes of RB convection}
\label{fig1}    
\end{figure}

The multi-layer, specifically bi-layer in our case, convection has traditionally been investigated to elucidate the liquid encapsulated crystal growth\cite{Johnson1975} and the Earth mantle convection \cite{Busse1981}. The liquid-liquid interaction in bi-layer engenders a plethora of thermo-convective modes. For a pair of immiscible fluids superposed in a horizontal configuration, \citet{Zeren1972} listed the ratios of the thermophysical properties of the fluid, the total height of the layer, and the height fraction of one of the layers as the key parameters that govern the convection patterns. Three distinct convection onset configurations, namely, buoyancy-dominated, interfacial tension gradient-dominated, and surface-deflection-dominated, may emanate based on the height fraction of the layer. For buoyancy-dominated configuration, the height fraction of the layer aids in further categorizing the convection onset on the basis of the relative propensity of the individual layers to undergo convection onset. Often, the buoyancy forces in the thicker layer leads to the flow onset in the layer, which subsequently drags the other layer owing to the viscous coupling at the interface \cite{Johnson1997}. Thus, one may get ``lower'' dragging (the dominant flow in the lower layer drags the upper layer) or the ``upper'' dragging mode. The experimental investigations \cite{Rasenat1989} for intermediate height fractions revealed the existence of the mechanical coupling (MC) mode and the thermal coupling (TC) mode. The MC mode is associated with the formation of two counter-rotating rolls, whereas the TC mode entails two co-rotating rolls. Figure \ref{fig1} depicts a schematic of the various flow configurations enumerated above. The bottom plate is at a higher temperature than the top plate $(T_B > T_T)$. The kinematic conditions at the interface are satisfied by the formation of intermediate eddy rolls in the case of TC. The competition between the MC and the TC modes at certain height fractions may lead to oscillatory onset of convection even if the interface has no distortion. Such oscillatory onset may be referred to as the ``oscillatory coupling instability'' \cite{Rasenat1989}. Alternatively, an interface admitting distortions may undergo ``oscillatory interfacial instability'' as a consequence of competition between localized destabilizing buoyancy effect and the stabilizing surface tension forces. The oscillatory modes gain more prominence in the solution spectrum when the distortion \cite{RenardyJoseph1985,Renardy1985} of the interface is allowed, typically observed in small density difference cases. If the surface tension gradient is strong enough, a surface tension-driven mode may arise, as shown in Fig.\ref{fig1}. Such a convection pattern is typically observed in liquid-gas systems \cite{Johnson1997}.  

\citet{Colinet1994} proposed the ``concept of balanced contrasts'' to explain the attainability of oscillatory mode of convection in buoyancy-dominated configuration. The Rayleigh number $(Ra)$ in the individual layers must be almost similar to have the prospect of oscillations. \citet{RenardyYY1996} found the oscillatory modes to occur when the ratio combination of the thermophysical properties namely density $(\rho)$, thermal expansion coefficient $(\beta)$ and the thermal diffusivity $(\kappa_T)$ is far from unity, i.e. $(\rho \beta \kappa_T)_r >> 1$.  \citet{Degen1998} experimentally substantiated the ratio combination criteria by considering the water-silicone oil system and further argued the importance of the height fraction of the lower layer. A switch between MC mode and TC mode is attainable only if sufficient thermal energy is transported to the interface to revert the flow direction in the upper layer. Most real-world systems, however, undergo the combined effect of buoyancy and thermo-capillarity \cite{Neopmnyashchy2004}. The inclusion of thermo-capillary flow renders the necessary criterion suggested by \citet{RenardyYY1996} insufficient. In our previous study \cite{Diwakar2014}, we presented the critical height ratio corresponding to equal Rayleigh number $Ra$ in both layers as a complimentary condition to the favourable thermophysical property combination. \\

Despite the continued interest, the binary fluids exhibiting temperature-sensitive miscibility gap is yet to be fully explored for various classical hydrodynamic instabilities. A modified phase field approach has been previously utilized to investigate the Rayleigh-Taylor  \cite{Lyubimova2019} and Kelvin-Helmholtz instabilities \cite{Zagvozkin2019,Kheniene2015}. In our previous studies \cite{Bestehorn2021,Borcia2022}, we reformulated the phase-field approach to incorporate an additional parameter characterizing the proximity of the system temperature to the UCST of the binary fluids. Recently, a linear analysis of Rayleigh-B\'enard-Marangoni (RBM) in general binary fluids has been conducted \cite{Mishra2025} based on the miscibility model suggested by Bestehorn et. al.\cite{Bestehorn2021}. However, the model \cite{Bestehorn2021} does not properly account for the Korteweg stresses while capturing the transition from the immiscible/partially miscible state to the miscible state. We proposed another model \cite{Borcia2022} to rectify the arbitrary reduction of the concentration gradient component. Nonetheless, the model fails to initialize the correct concentration profile in the domain. The erroneous concentration profile fails to model the partial miscibility between the two fluids. Thus in the current study, we employ an improved phase-field model which, on the one hand, correctly initializes the concentration profile and, on the other hand, leads to a miscibility-dependent surface tension. Employing the improved model, we analyze the convection onset criteria for a pair of binary fluids, specifically the FC72- 1cSt silicone oil combination. The system of fluids is considered at distinct temperatures to obtain the equilibrium concentration profiles at different degrees of partial miscibility. A linear stability analysis is performed by imposing a perturbation with infinitesimal amplitude. The dimensionless-linearized governing equations are obtained by employing the normal mode analysis. The resulting equations are discretized employing the Chebyshev pseudo-spectral method and subsequently solved to obtain the critical Rayleigh number $(Ra_c)$ and the growth rate of the imposed perturbation. One of the key objectives of this study is to reveal the window of wavenumbers of the perturbation that shall result in oscillatory convection. This window is analyzed as the two fluids are allowed to achieve a higher degree of partial miscibility. We further analyze the time period of these oscillations and explore the effects of the divergence of the interfacial thickness as the system temperature approaches the UCST.  \\ \\

The rest of the paper is organized as follows: Sec. \ref{sec:PFM} describes the phase-field model which is employed to reformulate the governing equations in Sec. \ref{sec:sysConfig}. Sec. \ref{sec:numericalFormulation} elucidates the numerical methodology employed to obtain the results discussed in Sec. \ref{sec:resultsDiscussion}. In Sec. \ref{sec:conclusion}, concluding remarks are provided.

\begin{figure*}
\centering 
\includegraphics[width=0.4\textwidth]{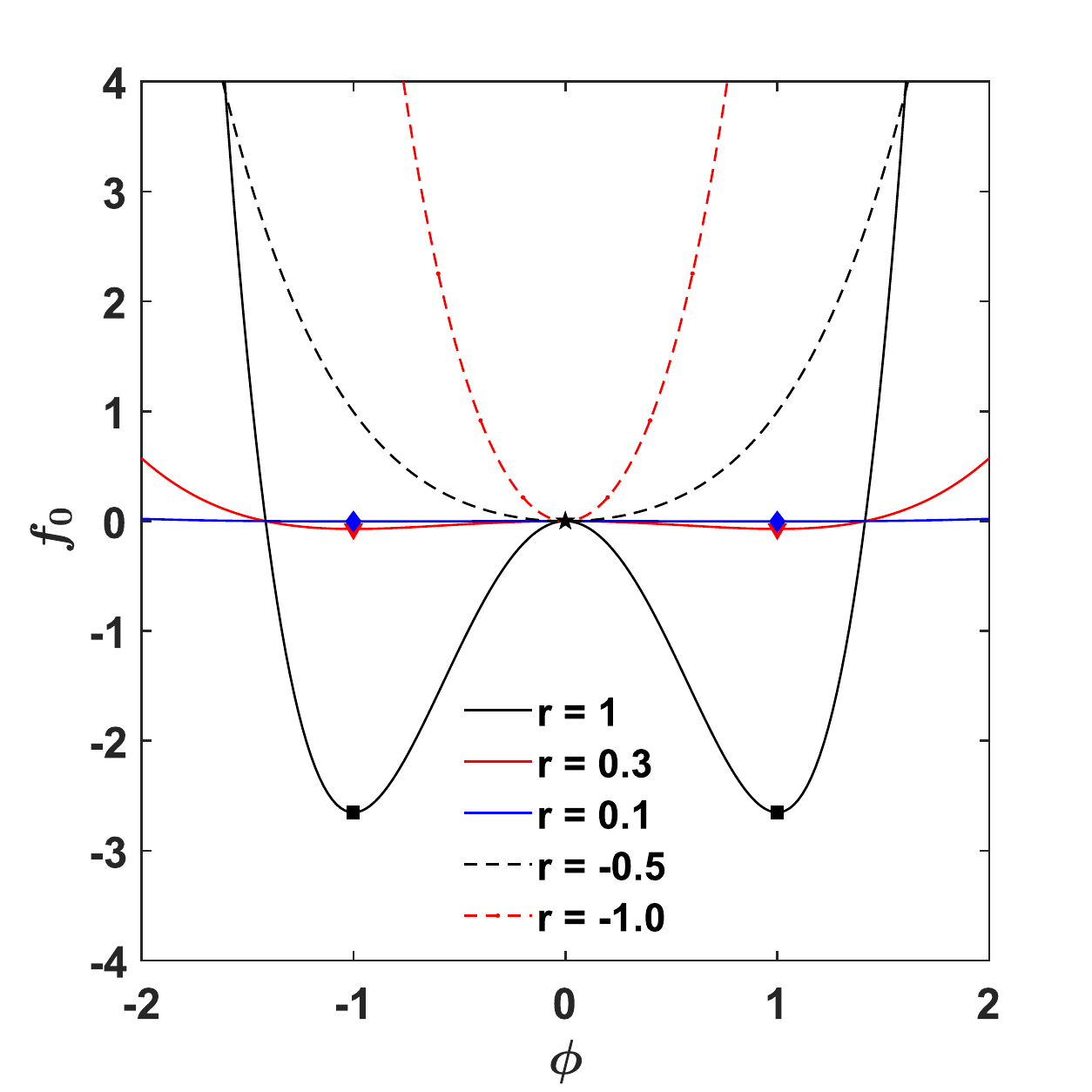} \hspace{0.2cm}
\includegraphics[width=0.4\textwidth]{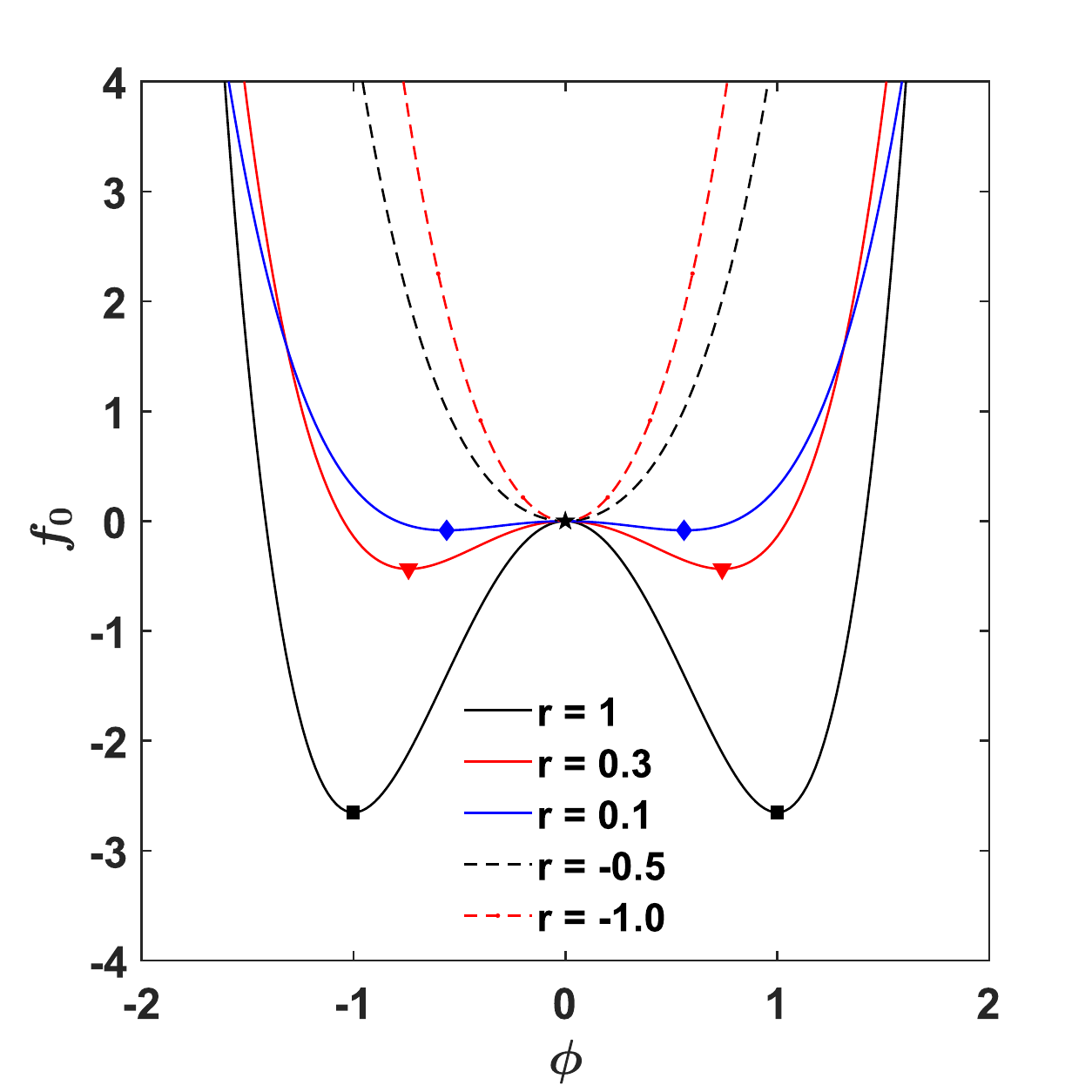} \\
\hspace{0.1cm} (a) \hspace{7.2cm} (b) \\
\caption{Bulk free energy transition from partially miscible state ($r>0$) to miscible state ($r<0$): $(a)$ Borcia et al. \cite{Borcia2022} and $(b)$ Current model}
\label{fig2}
\end{figure*}

\section{Phase-field model}
\label{sec:PFM}

The phase-field model \cite{Cahn1958} allows the description of a bi-layer problem within a single fluid formulation. The interface between the two fluids is modelled as a thin transition zone. The thermophysical properties vary rapidly but smoothly over this transition zone. The thickness of the transition zone, also known as interfacial thickness, is finite and is governed by the thermodynamic equilibrium state of the system. An order parameter, $\phi \in [-1,1]$, is defined to accommodate the spatio-temporal changes of an intensive variable due to inhomogeneity. A free energy functional that entails the contributions of bulk free energy from both the phases (liquids) and the excess energy ascribed to the presence of the interface is defined as a function of the order parameter. The free energy functional can be expressed as \cite{Yue2004,Ding2007}:   

\begin{equation}
 {F}(\phi, \bnabla \phi, T) = \int_{{\Omega}} [{f_{o}}(\phi) + {{\Lambda \over 2}}|{{\bnabla} \phi}|^2] d{\Omega}, 
 \label{freeEnergyFunct}
\end{equation}

where $f_{o}(\phi)$ is the bulk free energy density and $\Lambda$ is the mixing energy density. The mixing energy density is assumed to vary linearly with temperature $(\Lambda = \Lambda_0 - \Lambda_T T )$ to account for the thermo-capillarity. The bulk free energy determines the preferred state of the system in thermodynamic equilibrium, as shown in Fig. \ref{fig2}. If the bulk free energy has a form of double-well potential \cite{Ginzburg1950}, the system prefers to phase segregate into states that correspond to the minimum bulk free energy. Capitalizing on this feature, we reformulated the bulk free energy to capture the continuous transition of the binary fluids from the immiscible state to the miscible state in our previous studies \cite{Bestehorn2021, Borcia2022}. The modified expression of the bulk free energy incorporates an additional parameter, $r$, characterizing the proximity of the system temperature to the UCST. However, the previously proposed models do not accurately account for the Korteweg stresses \cite{Joseph1990} while the system is driven towards the state of miscibility \cite{Pojman2006}. In the first model \cite{Bestehorn2021} the arbitrary reduction of the concentration gradient component of the total free energy functional leads to under prediction of the surface tension forces. Subsequently, the Marangoni stress is also under-predicted. When a binary fluid system is allowed to attain the equilibrium state corresponding to different temperatures, the thermophysical properties and the interfacial thickness readjust accordingly to accommodate the revised distribution of fluid particles in the domain. The interfacial thickness diverges monotonically as we approach the UCST. The drawback of the first model was rectified in the second model \cite{Borcia2022}. However, the coefficients associated with the definition of the bulk free energy, i.e., with the latter defining the state of thermodynamic equilibrium, are given in an incorrect form. As a consequence, even when the system temperature approaches the UCST, the global minima of the bulk free energy curve occurs at $\phi = \pm 1$, as shown in Fig. \ref{fig2}(a). Thus, thermophysical properties of the two bulk phases, which are functions of the order parameter $\phi$, do not change, thereby rendering the model impractical and theoretically incorrect. Therefore, in this study, we employ an accurate phase-field model that couples the adequate features of the previously discussed models and, at the same time, rectifies the errors associated with them. Here, we define the bulk-free energy as follows:   

\begin{equation}
 {f_{0}}(\phi) = {{\Lambda} \over {\epsilon}^2}({1 \over 4}{|r|^a}{\phi^4} - {1 \over 2}r{\phi^2}), 
 \label{bulkFreeEnergy}
\end{equation}
\\
where ${\epsilon}$ is the interfacial thickness, with  

\begin{equation}
 r = f\bigg({{T - {T_c}} \over {T_{c}}}\biggr), 
 \label{misicbCoeff}
\end{equation}

${T_{c}}$ is the UCST, and ``$a$''  is an empirical coefficient (where $0< a < 1$) that has to be determined experimentally for a given pair of fluids. The limits on the value of the coefficient $a$ ensure the correct initialization of the concentration profile, as shown in Fig. \ref{fig2}(b). The dimensionless parameter $r$ is a monotonic function of the reduced system temperature $(\theta = {{T - {T_c}} \over {T_{c}}})$. If the temperature of the system is higher than the UCST, i.e. $r < 0$, the bulk free energy presumes a single well potential form and the system is driven towards the state of complete miscibility. On the other hand, if the system temperature is less than the UCST $(0 < r < 1)$, the two fluids can co-exist in an equilibrium state with a partially miscible configuration. The degree of partial miscibility is inversely proportional to the value of $r$. The miscibility parameter $r$ is defined as:  

\begin{equation}
 r = {e^{(-c\theta)} - e^{(c\theta)} \over e^{(-c\theta)} + b e^{(c\theta)} }, 
 \label{misicbCoeffExpression}
\end{equation}

\begin{figure}
\centering 
\includegraphics[width=0.45\textwidth]{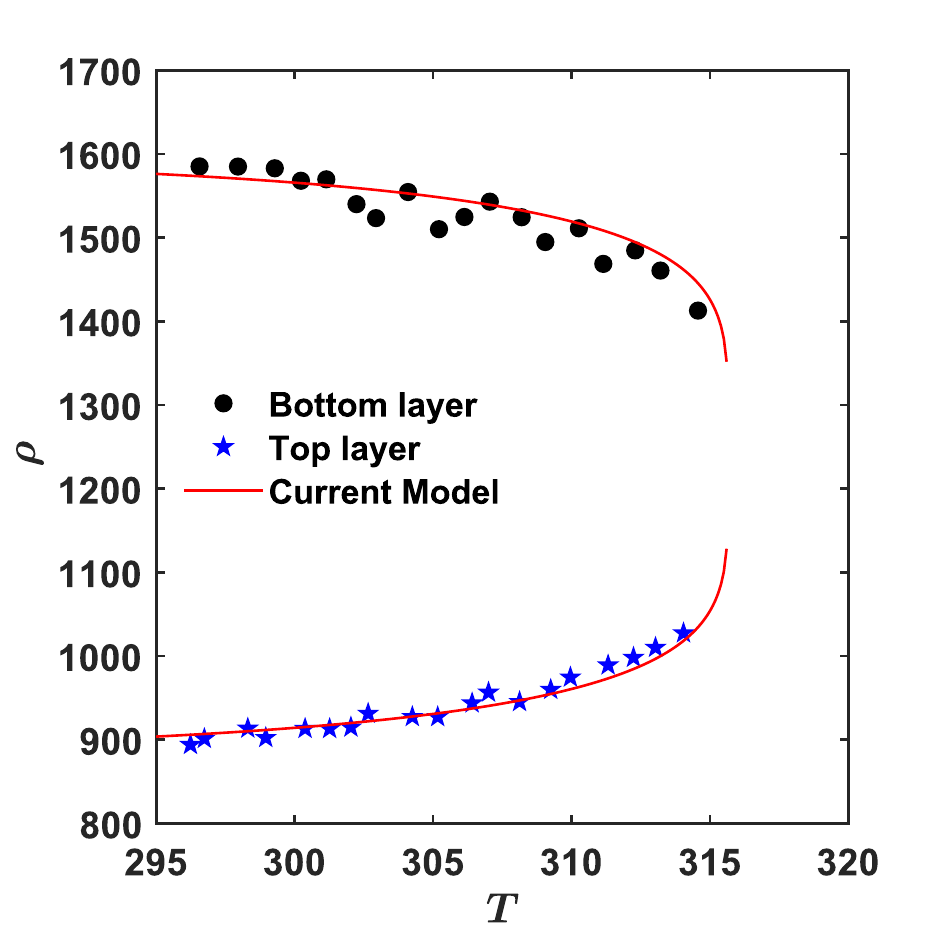} 
\caption{Reproducing experimentally observed variation of densities in FC72 rich region and silicone oil rich region as a function of the system temperature.}
\label{fig3}    
\end{figure}

\begin{figure*}
\centering 
\includegraphics[width=0.4\textwidth]{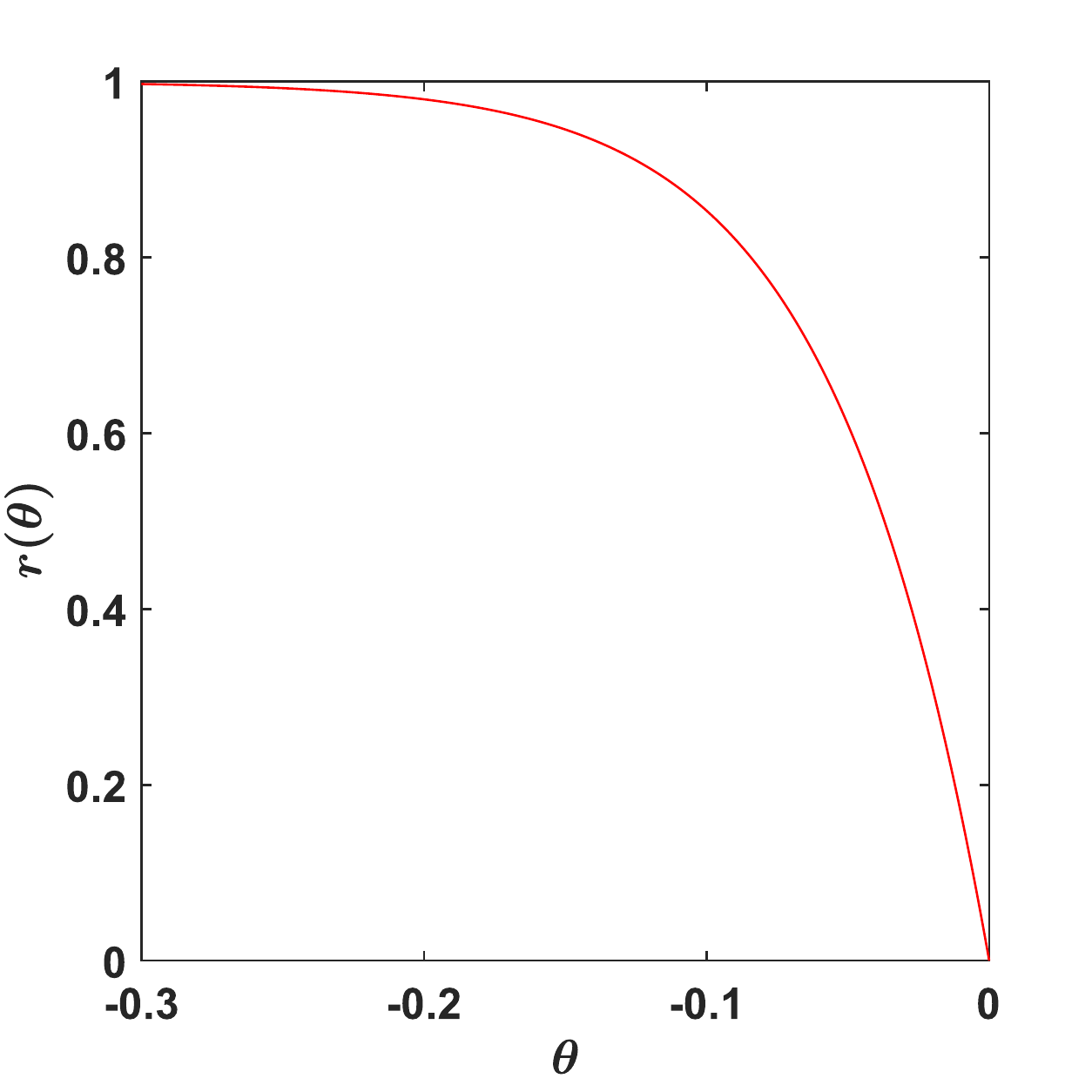} \hspace{0.2cm}
\includegraphics[width=0.4\textwidth]{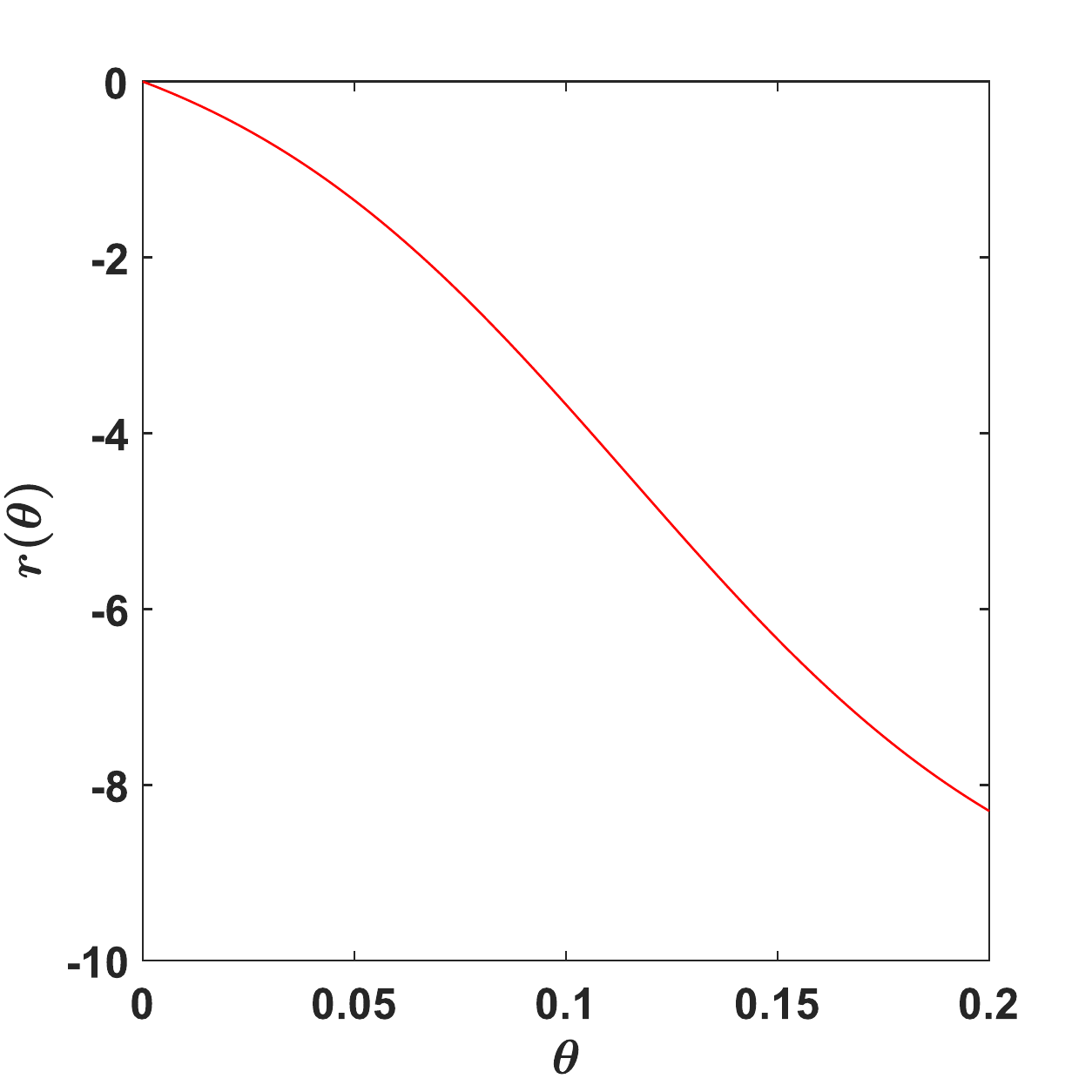} \\
\hspace{0.1cm} (a) \hspace{7.2cm} (b) \\
\caption{Variation of $r$ with $\theta$ for $c = 10$ and $b = 0.1$: $(a)$ $\theta < 0$ i.e. $T<T_c$  and $(b)$ $\theta > 0$ i.e.  $T > T_c$}
\label{fig4}
\end{figure*}

The empirical constants, $c$ and $a$, allow the model to accurately reproduce the density field and the surface tension variation as a function of temperature in the pre-UCST limit. For the present FC72 and silicone oil system, the constants required to obtain the correct density profile are $c = 10$ and $a = 0.6$. Figure \ref{fig3} demonstrates the ability of the model to reproduce the experimentally observed \cite{Jajoo2017} variation of densities in the FC72-rich region and the silicone oil-rich region. In the pre-UCST limit, the value of the constant $b$ has negligible effect. However, it plays a significant role when the system temperature is above the UCST, wherein it determines the rate of mass diffusion across the interface. It is to be noted that even when the system temperature is above UCST, the interface preserves its definition as long as the concentration gradient is non-trivial \cite{Pojman2006}. Figure \ref{fig4} gives an idea of how $r$ varies with $\theta$ for $c = 10$ and $b = 0.1$. The formulations allow the rate of diffusion to be a function of base system temperature when $\theta>0$.

The state of equilibrium is marked with zero chemical potential $(\mu = 0)$. The chemical potential, $\mu$, defined as the variational derivative of the total free energy functional, is expressed as   

\begin{equation}
\mu = {\delta F \over \delta \phi} =  \Lambda\Bigg(\frac{  r^a \phi^3 - r \phi} {\epsilon^2} -  \nabla^2 \phi \Bigg)
\label{chemicalPotential}
\end{equation} 

One can analytically derive the initial concentration profile and the surface tension associated with the diffuse interface between two fluids in a partially miscible configuration. The order parameter distribution and the surface tension are given respectively as

\begin{equation}
 \phi = - r^{(1-a) \over 2} \tanh\Bigl({{{y}-{y_{0}} \over {\sqrt{2} ({\epsilon} / \sqrt{r}) }}}\Bigr), 
 \label{equilibInterfaceProfile}
\end{equation}

and 

\begin{equation}
 {\sigma} = {{2 \sqrt{2}} \over 3} {{\Lambda} \over {\epsilon}} r^{(3-2a) \over 2} = {\sigma_0}r^{(3-2a) \over 2}.
 \label{surfaceTensionExpression}
\end{equation}

Here, ${y}_0$ is the location of the horizontal interface. It can be inferred from Eqns. \ref{equilibInterfaceProfile} and \ref{surfaceTensionExpression} in a straightforward manner that $r = 1$ reproduces the expressions for immiscible fluids \cite{Jacqmin1999, Yue2004}.

\section{System configuration and the governing equations}
\label{sec:sysConfig}

\begin{figure}
\centering 
\includegraphics[width=0.5\textwidth]{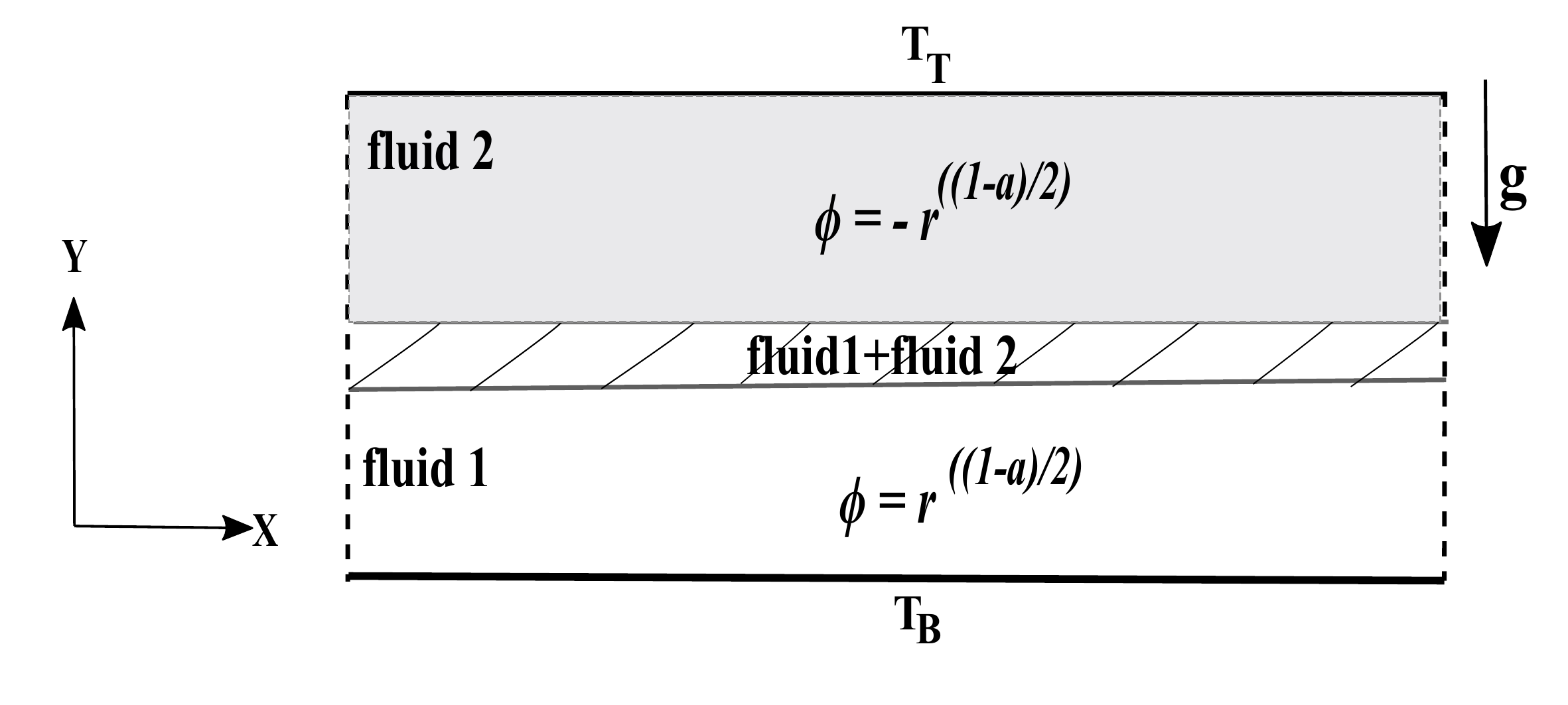} 
\caption{Schematic of pair of fluids heated from below}
\label{fig5}    
\end{figure}

Binary fluids exhibiting temperature-sensitive miscibility gaps are considered in a superposed configuration, as shown in Fig. \ref{fig5}. The lighter fluid (fluid 2) lies over the heavier fluid (fluid 1) to avoid the manifestation of the Rayleigh-Taylor instability. While the height of the domain is finite, it is infinite in the horizontal direction. The objective of the current study is to investigate the effect of distinct system temperatures on the onset of RBM convection in this bi-layer setup involving FC-72 and 1cSt silicone oil. The system is maintained at temperatures below UCST, i.e., $r>0$, throughout this work. The thermophysical properties of the fluid pair are given in table \ref{tab:table1}.       

\begin{table}[ht]
    \centering
    \caption{Thermophysical properties of FC-72 and Silicone}
    \label{tab:table1}
    \begin{ruledtabular}
    \begin{tabular}{lccccc}
        \textbf{Material} & $\rho$ (kg/m$^3$) & $\eta$ (kg/m·s) & $\beta$ (1/K) & $\kappa$ (W/m·K) & $C_p$ (J/kg·K) \\
        \hline
        FC-72  & 1680  & $6.38\times10^{-4}$ & 0.00156 & 0.057 & 1100  \\
        Silicone (Si)  & 816  & $1.0\times10^{-6}$ & 0.00134 & 0.1 & 2000  \\
    \end{tabular}
    \end{ruledtabular}
\end{table}

The two fluids are assumed to be Newtonian. The thermophysical properties of the bulk phases are assumed to be independent of temperature. For FC72-1cSt silicone oil, the UCST is $315.5K$. The surface tension at the interface between these two fluids\cite{Diwakar2018} far from UCST is $0.77 mN/m$. The miscibility parameter $r$ is assumed to be constant throughout the domain. This assumption is justified based on observations of \citet{Degen1998} as for a typical bi-layer setup, the critical temperature difference for the flow onset is of the order of 1$K$. The change in the miscibility parameter $r$ for such a low-temperature difference between the bottom and the top plate is meagre. The Navier-Stokes (NS) equations coupled with the continuity equation are employed to model the fluid flow. Further, the Boussinesq approximation, which allows the density field to be a function of the order parameter $\phi$ and the body force term to be a function of both the order parameter and the temperature $(T)$, is used. We deploy a volume-averaged velocity field \cite{Ding2007,Abels2013,Joseph1996} $\mathbf{u}$, allowing us to consider the divergence-free velocity condition. The NS equations (reformulated in terms of $r$) and the continuity equation are expressed as:

\begin{multline}
\rho (\phi) \Bigg(\frac{\partial \mathbf{u}}{\partial t} + \mathbf{u} \cdot \nabla \mathbf{u}\Bigg) = -\nabla p 
+ \nabla \cdot \left( \eta (\phi) \bigl[\nabla \mathbf{u} + (\nabla \mathbf{u})^T \bigr]\right) - \rho' g_{\vec{\mathbf{y}}} \\
+ \mu \nabla \phi 
+ \Bigl(\frac{1}{4 \epsilon^2} r^a \phi^4 - \frac{1}{2 \epsilon^2} r \phi^2 
+ \frac{1}{2}  |\nabla \phi|^2 \Bigl)\frac{\partial \Lambda}{\partial T} \nabla T \\
-  \nabla \phi \frac{\partial \Lambda}{\partial T} \nabla T \cdot \nabla \phi.
\label{NSEquation}
\end{multline}

\begin{equation}
\nabla \cdot \mathbf{u} = 0.
\label{ContinuityEquation}
\end{equation}

where   $\rho (\phi)$ is the average density field, $p$ is the pressure field, $\eta (\phi)$ is the average dynamic viscosity, $\rho' = \rho(\phi,\theta) - \rho_0$ with $\rho_0 = {\rho_1 + \rho_2 \over 2}$ and $\rho(\phi, T) = \rho_{1}(1 - \beta_{1}(T - T_{0}))({1 + \phi \over 2}) + \rho_{2}(1 - \beta_{2}(T - T_{0}))({1 - \phi \over 2})$.  

The spatiotemporal evolution of the interface, depicted by the order parameter $\phi$, coupled with the hydrodynamic flow, is governed by the Cahn-Hilliard equation \cite{Cahn1958,Jacqmin1999} as shown below.   

\begin{multline}
\frac{\partial \phi}{\partial t} + \mathbf{u} \cdot \nabla \phi = \nabla \cdot \Bigg( \gamma \nabla \Bigg\{\Lambda\Bigg[\frac{  r^a \phi^3 - r \phi} {\epsilon^2} -  \nabla^2 \phi \Bigg]\Bigg\} \Bigg)
\label{CahnHilliardEquation}
\end{multline} 

\noindent where $\gamma$ is the mobility of the diffuse interface. An optimum value of mobility is required to avoid, on the one hand, the shear-thinning of the interface, while on the other hand, the over-damping of the flow induced due to high diffusion \cite{Magaletti2013}. Finally, the energy equation \cite{Mishra2025} is given as 

\begin{equation}
\rho c \Bigg(\frac{\partial T}{\partial t} + \mathbf{u} \cdot \nabla T \Bigg)
+ \Lambda_0 \Big(\frac{r^a \phi^3 - r \phi }{\epsilon^2} -  \nabla^2 \phi \Big) \frac{D\phi}{Dt} 
= \nabla \cdot (\kappa \nabla T),
\label{energyEquation}
\end{equation}

\noindent where $\kappa$ is the thermal conductivity, $c$ is the specific heat. The thermophysical properties can be expressed as a function of order parameter as 

\begin{align}
   \frac{\rho}  {\rho_2} &= \frac{1}{2} (1 + \frac{\rho_1}  {\rho_2}) +\frac{\phi}{2}  (\frac{\rho_1}  {\rho_2} - 1) \notag \\
   \frac{\eta}  {\eta_2} &= \frac{1}{2} (1 + \frac{\eta_1}  {\eta_2}) +\frac{\phi}{2}  (\frac{\eta_1}  {\eta_2} - 1) \notag \\
    \frac{\kappa} {\kappa_2} &= \frac{1}{2} (1 + \frac{\kappa_1} {\kappa_2}) + \frac{\phi}{2}  (\frac{\kappa_1} {\kappa_2} - 1) \notag \\
    \frac{\rho c} {\rho_2 c_2 }&= \frac{1}{2} (1 + \frac{\rho_1 c_1 }{\rho_2 c_2}) + \frac{\phi}{2} (\frac{\rho_1 c_1 }{\rho_2 c_2} - 1) \notag \\
    \frac{\kappa_T} {\kappa_{T2}} &= \frac{1}{2} (1 + \frac{\kappa_{T_{1}}} {\kappa_{T_{2}}}) + \frac{\phi}{2} (\frac{\kappa_{T_{1}}} {\kappa_{T_{2}}} - 1)
    \label{thermophysicalProperties}
\end{align}

where $\kappa_T$ is the thermal diffusivity of the fluid. 

\subsection{Linearized Governing equations}
\label{subsec:LinearGDE}

In this subsection, we provide the details pertinent to the non-dimensionalization of the governing equations. The reference quantities (denoted by subscript ``R'') used for primary flow variables are based on fluid $2$ properties as follows. 

\begin{align}
L_R = H_2  \qquad t_R = {{H_2}^2 \over {\kappa_T}_2} \qquad u_R = {{\kappa_T}_2 \over H_2} \qquad \nonumber \\
T_R = T_B - T_T \qquad  p_R = \rho_2 \bigl({{\kappa_T}_2 \over H_2}\bigr) \bigl({\nu_2  \over {H_2}}\bigr)
\label{referenceQuantities}
\end{align}

The flow variables are perturbed with an infinitesimal disturbance for the linear analysis to be valid. The perturbed variables can be expressed as: 

\begin{align}
\phi &= \phi^b(y) + \phi'(x,y,t), \\
 u_i &= u_i'(x,y,t), \\
T &= T^b(y) + T'(x,y,t), \\
p &= p^b(y) + p'(x,y,t).
\label{perturbedVariables}
\end{align}

where the superscript ``b'' indicates the base quantities prior to perturbation and ``$ \prime $'' indicates the imposed disturbance. We further assume the disturbance $X = \{\phi', u', v', p', T'\}$ to follow normal mode analysis, wherein they can be considered in periodic form as follows.

\begin{align}
X = \hat{X}(y)e^{(\lambda t + \iota k x)}, 
\label{perturbationNormalDecomposition}
\end{align}

\noindent where $ \lambda $ is the growth rate and $k$ is the wavenumber of the perturbation. Substituting the normal mode decomposed values into the governing equations (Eqns. \ref{NSEquation}, \ref{ContinuityEquation} and \ref{CahnHilliardEquation}) and linearizing, we obtain the following dimensionless equations for the perturbed quantities: 
\\
\\
Continuity:
\begin{align}
    i k \hat{u} + \frac{d\hat{v}}{dy} = 0,
\label{linNonDim_continuityEquation}
\end{align}
\\
x-momentum:
\begin{align}
    \lambda \hat{u} &=  
    - \frac{Pr_2  (i k) }{\rho_f(\Phi^b)} \hat{p} 
    + \frac{Pr_2 }{\rho_f(\Phi^b)} 
    \Bigg[ \frac{1}{2} (1 + \frac{\eta_1} {\eta_2}) 
    + \frac{\Phi^b}{2} (\frac{\eta_1} {\eta_2} -1) \Bigg] \notag \\ 
    &\quad \Bigg( -k^2 \hat{u} + \frac{d^2 \hat{u}}{dy^2} \Bigg) \notag \\
    &\quad + \frac{Pr_2 }{2 \rho_f(\Phi^b)} (\frac{\eta_1} {\eta_2} -1) 
    \Bigg( \frac{d\Phi^b}{dy} \frac{d\hat{u}}{dy} 
    + \frac{d\Phi^b}{dy} (ik\hat{v})) \Bigg) \notag \\
    &\quad + \frac{Pr_2  Ra_2 }{\rho_f(\Phi^b)} \frac{3}{2\sqrt{2}} Y  
    \frac{\varepsilon }{H_2} 
    \Bigg[ 
     \frac{1}{2}
    \Bigg(\frac{d \Phi^b}{ dy}\Bigg)^2 \Bigg] i k \hat{T} \notag \\
    &\quad - \frac{Pr_2  Ra_2 }{\rho_f(\Phi^b)} \frac{3}{2\sqrt{2}} Y 
    \frac{\varepsilon}{H_2 }  
    \Bigg( \frac{d \Phi^b}{d y} \frac{d T^b}{d y} \Bigg) i k \hat{\phi}.
\label{X_momentumEquation}
\end{align}
Cahn-Hilliard: 
\begin{multline}
    \lambda \hat{\phi} + \hat{v} \frac{\partial \Phi^b}{\partial y} =
    M \Bigg[  3r^a 
        \Bigg(- (\Phi^b)^2 k^2 \hat{\phi} 
        + 2 \left(\frac{d \Phi^b}{dy}\right)^2 \hat{\phi} \\
        + 2 \Phi^b \frac{d^2 \Phi^b}{dy^2} \hat{\phi} 
        + 4 \Phi^b \frac{d \Phi^b}{dy} \frac{d \hat{\phi}}{dy} 
        + (\Phi^b)^2 \frac{d^2 \hat{\phi}}{dy^2} \Bigg)
     \\
    - r  (- k^2 \hat{\phi} 
    + \frac{d^2 \hat{\phi}}{dy^2} )
    - (\frac{\varepsilon} {H_2 })^2  \bigg(\frac{d^4 \hat{\phi}}{dy^4} \\
    - 2 k^2 \frac{d^2 \hat{\phi}}{dy^2} 
    + k^4 \hat{\phi} \bigg) \Bigg]\\
    - M \frac{Y Ra_2}{Ca} T^b \Bigg\{ 
        3 r^a \Big[ 
            - (\Phi^b)^2 k^2 \hat{\phi} 
            + 2 \left(\frac{d \Phi^b}{dy}\right)^2 \hat{\phi} \\
            + 2 \Phi^b \frac{d^2 \Phi^b}{dy^2} \hat{\phi} 
            + 4 \Phi^b \frac{d \Phi^b}{dy} \frac{d \hat{\phi}}{dy} 
            + (\Phi^b)^2 \frac{d^2 \hat{\phi}}{dy^2} 
        \Big] \\
        - r (- k^2 \hat{\phi} 
        + \frac{d^2 \hat{\phi}}{dy^2} )
        - (\frac{\varepsilon}{H_2 })^2 \bigg(\frac{d^4 \hat{\phi}}{dy^4} \\
        - 2 k^2 \frac{d^2 \hat{\phi}}{dy^2} 
        + k^4 \hat{\phi}\bigg) 
    \Bigg\} \\
    - 2 M \frac{Y Ra_2}{Ca} \frac{d T^b}{dy} \Bigg[
        3 r^a (\Phi^b)^2 \frac{d \hat{\phi}}{dy} 
        + 6 r^a (\Phi^b) \frac{d \Phi^b}{dy} \hat{\phi} \\
        - r \frac{d \hat{\phi}}{dy} 
        - (\frac{\varepsilon}{H_2 })^2 \bigg( \frac{d^2 }{dy^2} 
        - k^2  \bigg) \frac{d \hat{\phi}}{dy}
    \Bigg] \\
    - M \frac{Y Ra_2}{Ca} \frac{d^2 T^b}{dy^2} \Bigg[ 
        3 r^a (\Phi^b)^2 \hat{\phi} 
        - r \hat{\phi} 
        - (\frac{\varepsilon}{H_2 })^2   \nabla^2 \hat{\phi} 
    \Bigg],
    \label{linNonDim_CHEquation}
\end{multline}
\\
\\
\\
Energy:
\begin{align}
    (\rho c)_f(\Phi^b) \Bigg[ \lambda \hat{T} + \hat{v} \frac{\partial T^b}{\partial y}\Bigg] &=  
    \left( \frac{1}{2} (1 + \frac{\kappa_1} {\kappa_2}) + \frac{\Phi^b}{2} (\frac{\kappa_1} {\kappa_2} - 1) \right) \notag \\
    &\quad \left( -k^2 + \frac{d^2}{dy^2} \right) \hat{T} \notag \\
    &\quad + \frac{1}{2} (\frac{\kappa_1} {\kappa_2}- 1) \frac{d \Phi^b}{dy} \frac{d \hat{T}}{dy} \notag \\
    &\quad + \frac{1}{2} (\frac{\kappa_1} {\kappa_2} - 1) \frac{d \hat{\phi}}{dy} \frac{d T^b}{dy} \notag \\
    &\quad + \Bigg[\frac{\hat{\phi}}{2} (\frac{\kappa_1} {\kappa_2} - 1) \Bigg]\frac{d^2 T^b}{dy^2}.
    \label{linNonDim_energyEquation}
\end{align}
\\

y-momentum:
\begin{align}
    {\lambda}\hat v &= -\frac{\Pr_2}{\rho_f(\Phi^b)  } \frac{d\hat{p}}{dy} 
    + \frac{\Pr_2}{\rho_f(\Phi^b)  } \Bigg(\frac{1}{2} (1 + \frac{\eta_1}{ \eta_2} \notag \\
    &\quad + \frac{\Phi^b}{2} (\frac{\eta_1}{ \eta_2} - 1) \Bigg)(-k^2 + \frac{d^2}{dy^2}) \hat{v}  \notag \\
    &\quad + \frac{\Pr_2}{\rho_f(\Phi^b)  } (\frac{\eta_1}{ \eta_2} - 1) \frac{d\Phi^b}{dy} \frac{d\hat{v}}{dy} \notag \\
    &\quad - \frac{1}{\rho_f(\Phi^b)} \Gamma_T \frac{d\Phi^b}{dy} \Big( -k^2 + \frac{d^2}{dy^2}  \Big)\hat{\phi} \notag \\
    &\quad + \frac{1}{\rho_f(\Phi^b)} Ra_2  Pr_2  \frac{1}{2} \left( \frac{\rho_1 \beta_1}{\rho_2 \beta_2} -1 \right) T^b \hat{\phi}  \notag \\
    &\quad + \frac{1}{\rho_f(\Phi^b)} Ra_2  Pr_2 \Bigg[ \frac{1}{2} \left( \frac{\rho_1 \beta_1}{\rho_2 \beta_2} +1 \right) \notag \\
    &\quad + \frac{1}{2} \left( \frac{\rho_1 \beta_1}{\rho_2 \beta_2} -1 \right) \Phi^b \Bigg]\hat{T}  \notag \\
    &\quad + \frac{1}{\rho_f(\Phi^b)} \frac{3}{2\sqrt{2}} Y Pr_2  Ra_2  \frac{\varepsilon }{H_2} \big(\frac{1}{2}\big)   
    \left(\frac{d\Phi^b}{dy}\right)^2 \frac{d\hat{T}}{dy}  \notag \\
    &\quad + \frac{1}{\rho_f(\Phi^b)} \frac{3}{2\sqrt{2}} Y Pr_2  Ra_2  \frac{\varepsilon }{H_2} \Bigg(
    \frac{d\Phi^b}{dy} 
    \frac{d{T^b}}{d y} \Bigg)\frac{d \hat{\phi}}{d y} \notag \\
    &\quad + \frac{1}{\rho_f(\Phi^b)} \frac{3}{2\sqrt{2}} Y Pr_2 Ra_2 \frac{\varepsilon}{H_2 }   
    \frac{d \Phi^b}{d y} T^b (-k^2 + \frac{d^2}{dy^2}) \hat{\phi} \notag \\
    &\quad  - \frac{1}{\rho_f(\Phi^b)} Ga_2 \biggl[{1 \over 2} \bigl({\rho_1 \over \rho_2} -1 \bigr)\biggr] \hat{\phi}
    \label{Y_momentumEquation}
\end{align}

The process of non-dimensionalization leads to several dimensionless parameters, namely, the dimensionless density $\bigl(\rho_f(\Phi^b)  = {\rho(\Phi^b) \over \rho_2}\bigr)$, the dimensionless mobility ($M = {\gamma \Lambda_0 \over \epsilon^2 {\kappa_T}_2}$), the inverse Capillary number $(Ca = {\sigma H_2 \over \eta_2 {\kappa_T}_2})$, the Prandtl number $(Pr_2 = {\eta_2 c_2 \over  \kappa_2})$, the Rayleigh number $(Ra_2 = {\rho_2 g \beta_2 (T_B - T_T) {H_2}^3 \over \eta_2 {\kappa_T}_2})$ , the Marangoni number $(Ma_2 = {\sigma_T (T_B - T_T) H_2 \over \eta_2 {\kappa_T}_2})$ and $\Gamma_T (= {\Lambda_0 \over \rho_2 {{\kappa_T}_2}^2})$ is the modified inverse capillary number. The term `modified' signifies the usage of thermal diffusivity instead of kinematic viscosity. The geometric features of the domain are governed by the Galileo number $(Ga_2 = {{H_2}^3 g \over {{\kappa_T}_2}^2})$. Finally, we define a modified Bond number $(Y = {Ma \over Ra})$ characterizing the ratio of buoyancy forces due to gravity to the surface tension forces. 

\subsection{Initial and Boundary Conditions}
\label{subsec:Initial_BC}

Initially, the two fluids are assumed to be stationary. Consequently, the temperature profile in the domain is linear owing to pure conductive heat transport from the bottom plate to the top plate. The system is considered at distinct temperatures with varying degrees of proximity to the UCST. The interface profile is initialized employing equilibrium profile (eq. \ref{equilibInterfaceProfile}). The domain extends limitlessly in the horizontal direction. However, the domain is bounded in the vertical direction by impermeable isothermal walls. Consequently, the perturbation on the base flow satisfies the no-slip and no-penetration boundary conditions. Furthermore, the perturbation on the temperature profile is zero at the walls as the temperature is pre-specified. As the focus of the current study is on the onset of convection, the interfacial deformation is meagre. The mass flux through the top and bottom walls is also zero $(\mu = 0)$. Thus, in the dimensionless framework, we get the following boundary conditions:       

\begin{align}
    \frac{\partial \hat{\phi}}{\partial y} &= 0, \quad \frac{\partial^3 \hat{\phi}}{\partial y^3} = 0 \quad \text{at } y = 0,1 \\
    \hat{u}_i &= 0 \quad \text{at } y = 0,1 \\
    \hat{T} &= 0 \quad \text{at } y = 0,1
    \label{BoundaryConditions}
\end{align}

\section{Numerical Methodology}
\label{sec:numericalFormulation}

The prediction of critical values for the onset of convection imposes stringent accuracy requirements. Thus, the Chebyshev pseudo-spectral method is employed for spatial discretization of the pertinent variables \cite{Diwakar2014}. The variables can be expressed using Lagrange polynomial as follows.

\begin{align}
    \Xi_N(y) &= \sum_{i=0}^{N} h_i(y) \Xi(y_i),
    \label{LagrangePolynomial}
\end{align}

where $N$ is the number of collocation points in the vertical direction. The cardinal function $h_i(y)$ is given as:

\begin{align}
    h_i(y) &= (-1)^{i+1} (1 - y^2) \frac{T_N'(y)}{c_i N^2 (y - y_i)}
    \label{CardinalFunction}
\end{align}

with $c_i = 2$ for the endpoints and $1$ for interior points and $T_N'(y)$ is the $N^{th}$ derivative of the Chebyshev polynomial of the first kind. \\
The Gauss-Lobatto-Chebyshev (GLC) formulation leads to the concentration of the grid points close to the extremities or the domain boundaries. However, for the current problem, we need to concentrate the points around the diffuse interface to account for the Korteweg stresses accurately. Thus, we transform the grid points following the methodology suggested by \citet{Tee2006} and \citet{Diwakar2015}. The grid points are relocated using the following expression:

\begin{multline}
  y = g(y) = \delta  \\ + \alpha \sinh \Bigl[ \bigg(\sinh^{-1} \Bigl(\frac{1 - \delta}{\alpha} \Bigr) + \sinh^{-1} \Bigl(\frac{1 + \delta}{\alpha} \Bigr) \bigg) \bigg. 
  \bigg. \frac{y - 1}{2} \\
  + \sinh^{-1} \Bigl( \frac{1 - \delta}{\alpha} \Bigr)  \Bigr]   
  \label{Tee_Trefethen_Transformation}
\end{multline}

where $\delta$ is the location of the diffuse interface and $\alpha$ determines the concentration of the grid points around the interface. An optimum value of $\alpha (= 0.0001)$ is chosen so that the remaining section of the domain is not scarce of the grid points to resolve the flow variables. Subsequently, the spatial derivatives are calculated in Eqns. \ref{linNonDim_continuityEquation} - \ref{Y_momentumEquation} are accordingly modified to account for the relocation of the grid points \cite{Tee2006} as 

\begin{align}
    \frac{d \Xi}{d \tilde{y}} &= \frac{1}{g'} \frac{d \Xi}{d y} \\
    \frac{d^2 \Xi}{d \tilde{y}^2} &= \frac{1}{(g')^2} \frac{d^2 \Xi}{d y^2} - \frac{g''}{(g')^3} \frac{d \Xi}{d y} \\
    \frac{d^3 \Xi}{d \tilde{y}^3} &= \frac{1}{(g')^3} \frac{d^3 \Xi}{d y^3} - \frac{3 g''}{(g')^4} \frac{d^2 \Xi}{d y^2} - \frac{g' g'''}{(g')^5} - \frac{3 (g'')^2}{(g')^5} \frac{d \Xi}{d y} \\
    \frac{d^4 \Xi}{d \tilde{y}^4} &= \frac{1}{(g')^4} \frac{d^4 \Xi}{d y^4} - \frac{6 g''}{(g')^5} \frac{d3 \Xi}{d y^3} - \frac{4 g' g'''}{(g')^6} - \frac{15 (g'')^2}{(g')^6} \frac{d^2 \Xi}{d y^2} \notag \\
    &\quad - \frac{(g')^2 g''''}{(g')^7} - \frac{10 g' g'' g'''}{(g')^7} + \frac{15 (g'')^3}{(g')^7} \frac{d \Xi}{d y}
    \label{spatialDerivativeCalculation}
\end{align}

\subsection{Calculation of critical Rayleigh number}
\label{subsec:CalccriticalRayleighNumber}

The critical Rayleigh number corresponds to the state of marginal stability that is characterized by zero growth rate ($\lambda = 0$). The governing equations (Eqns. \ref{linNonDim_continuityEquation} - \ref{Y_momentumEquation}) are rewritten in the form $Ra_2 A_1 \underline{X} = B_1 \underline{X}$, where $\underline{X}$ is the vector containing unknowns, $A$ and $B$ are coefficient matrices. The eigenvalues for the set of equations are obtained by deploying the inbuilt eigenvalue calculator in Octave 6.4.0. The smallest eigenvalue becomes the critical Rayleigh number $(Ra_c)$.   

\subsection{Calculation of growth rate}
\label{subsec:CalcGrowthRate}

Having calculated the $Ra_c$, the equations can be rewritten in the form $\lambda A_2 \underline{X} = B_2 \underline{X}$ so that the growth rate $\lambda$ is the eigenvalue. The system of equation is solved in an analogous manner to obtain $\lambda = \lambda_r + \iota \lambda_{I}$ where $\lambda_r$ is the real component and $\lambda_{I}$ is the imaginary component of the growth rate. The temporal growth of the amplitude of the perturbation is governed by $\lambda_r$, whereas the prospect of oscillation is governed through $\lambda_{I}$, i.e., $\lambda_{I} \neq 0$

\begin{table*}[ht]
    \centering
    \caption{Dimensionless parameters required for validation and results }
    \label{tab:table2}
    \begin{ruledtabular}
    \begin{tabular}{cccccccccc}
       \textbf{Fluid1-Fluid2} & $\bm{\frac{\rho_1}{\rho_2}}$ & $\bm{\frac{\eta_1}{\eta_2}}$  & $\bm{\frac{\beta_1}{\beta_2}}$  & $\bm{\frac{\kappa_1}{\kappa_2}}$  & $\bm{\frac{C_1}{C_2}}$ &$\bm{Pr_2}$ &  $\bm{Ga_2}$ & $\bm{M}$ &$\bm{\Gamma_T}$  \\
        \hline
        Water-Si & 1.15  & 0.57 & 0.17 & 5.44 & 2.56 & 25.71& $10^9$& $7.15 \times  10^5$ & $1.06 \times  10^7$ \\
        FC72-1cSt Si & 2.06  & 0.78 & 1.16 & 0.57 & 0.55 & 16.32& $10^9$& 15000& $2.66 \times  10^6$\\
    \end{tabular}
    \end{ruledtabular}
\end{table*}

\section{Results and Discussion}
\label{sec:resultsDiscussion}

\begin{figure}
\centering 
\includegraphics[width=0.45\textwidth]{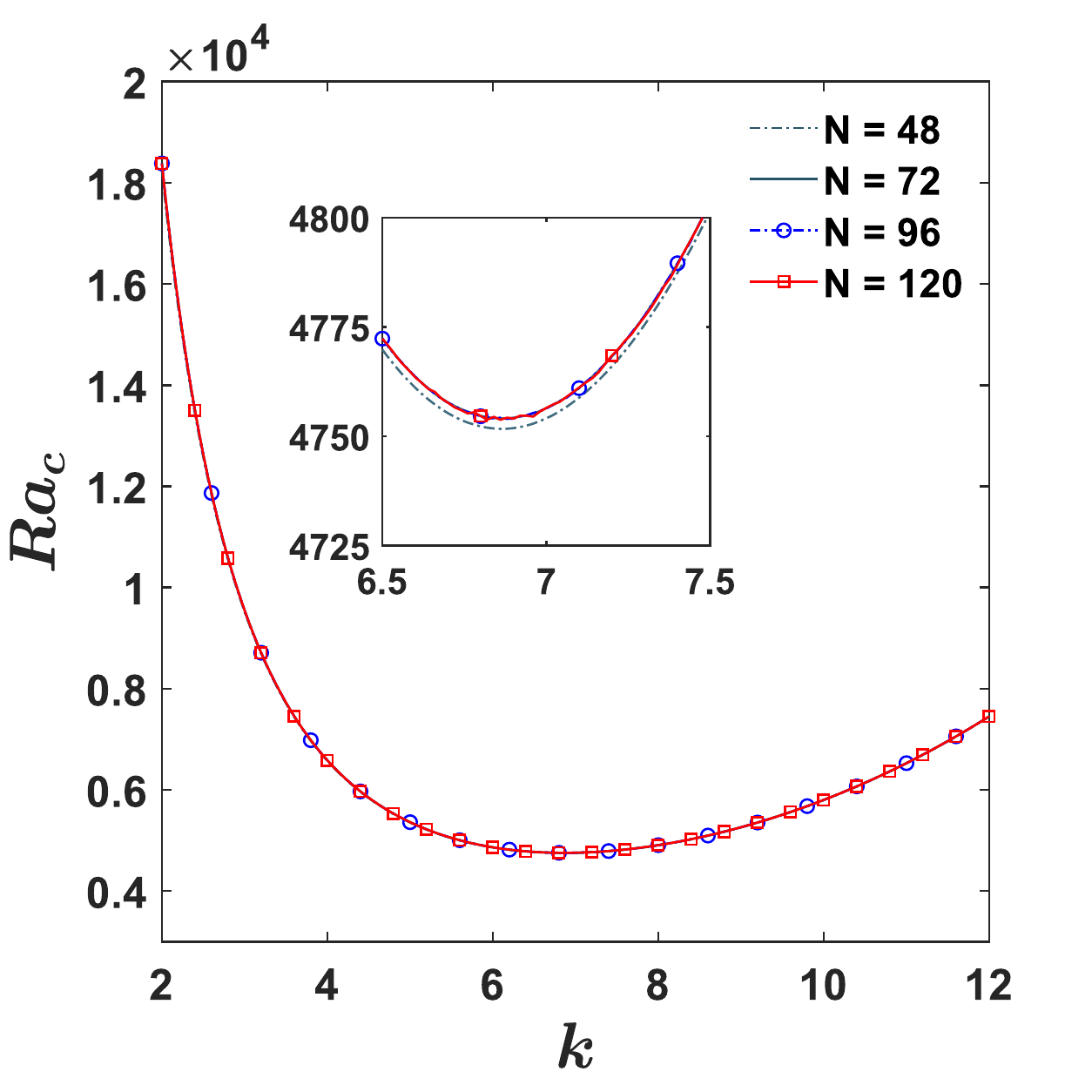} 
\caption{The variation of critical Rayleigh-Number as a function of wavenumber for four different grid resolutions namely $(a) N = 48$, $(b) N = 72$, $(c) N = 96$ and $(d) N = 120$. The inset in the figure further highlights the identical solution obtained for finer grids.}
\label{fig6}    
\end{figure}

\begin{figure}
\centering 
\includegraphics[width=0.45\textwidth]{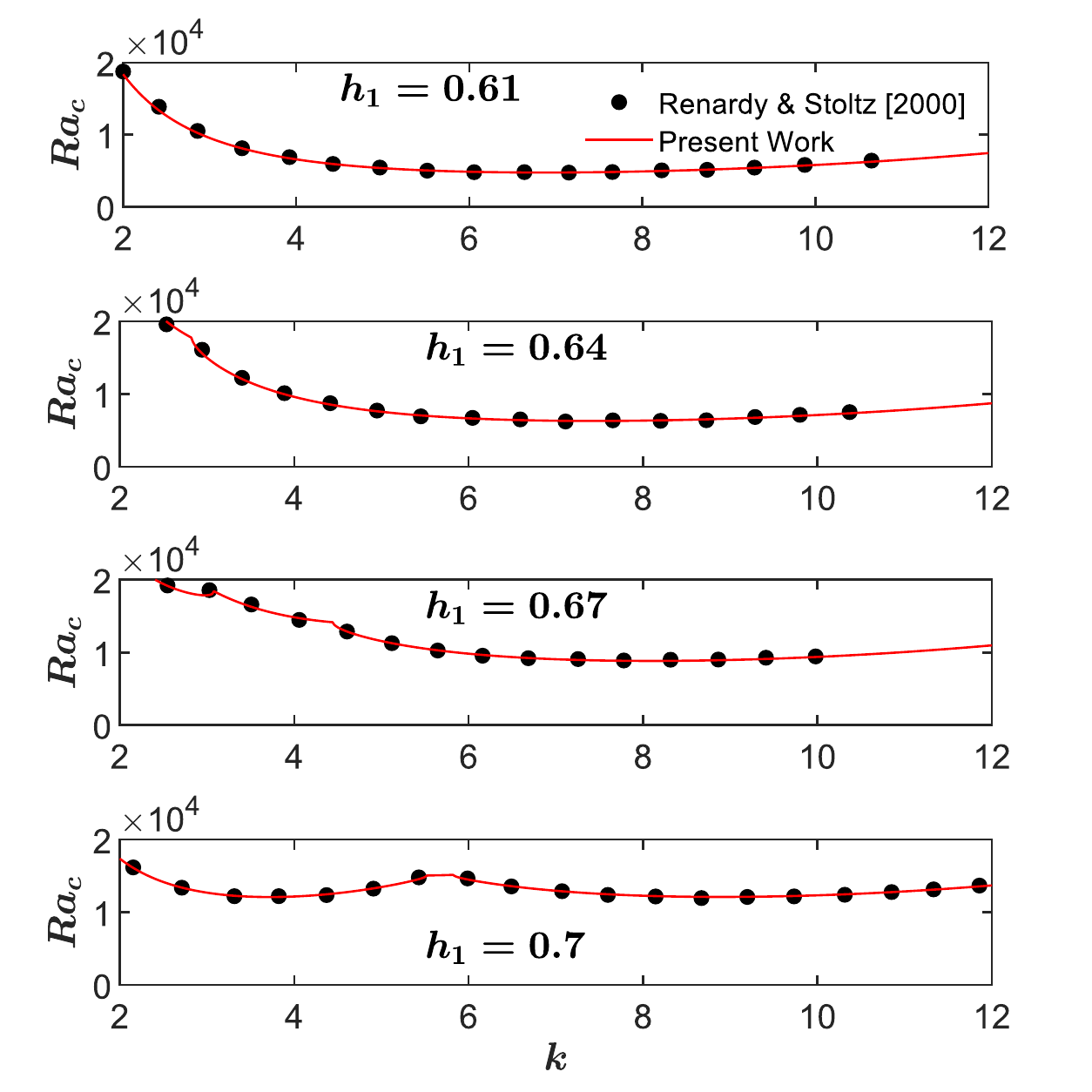} 
\caption{Validation of the present solver with the work of \citet{Renardy2000} for the water-silicone oil system.}
\label{fig7}    
\end{figure}

In this section, we discuss the results of the linear stability analysis. To begin with, we present a qualitative and quantitative validation of our solver. The values of the dimensionless parameters required to obtain the forthcoming results are given in table \ref{tab:table2}. A grid independence test is conducted to obtain the optimum number of points required for the stability calculations of a pair of binary fluids at the limit of immiscibility, i.e., $r \to 1$. Fig. \ref{fig6} depicts the variation of the $Ra_c$ as a function of the wavenumber $k$. We found that the $Ra_c$ and the most dangerous wavenumber are exactly the same for $N = 72$, $96$ and $120$. Thus, to minimize the computational time without loosing the accuracy of results, $N = 72$ grid points have been chosen here to generate all the forthcoming results. Our numerical solver is validated against the benchmark linear analysis results of \citet{Renardy2000} for an immiscible water-silicone oil system. The water-silicone oil system has a density ratio of 1.149. Therefore, this choice of fluid pairs for the validation demonstrates the ability of our solver to accurately predict the onset of convection in relatively low-density ratio fluid pairs that one might encounter in binary fluids exhibiting temperature-sensitive miscibility gaps. The $Ra_c$ variation with wavenumber $k$ for multiple height ratios corroborates the findings of \citet{Renardy2000}  as shown in Fig. \ref{fig7}. For $h_1 = 0.61 \to 0.69$, the onset mode lies for wavenumber beyond $5$, while for $h_1 = 0.7$, a switch in the tendency of provocation of convection is observed for a wavenumber less than $5$. This switch is attributed to the development of two lobes in the presence of Hopf modes \cite{Renardy2000}. Further, with $h_1 = 0.7$, our solver predicts the instability of oscillatory modes for wavenumbers ranging between $5.54$ and $5.82$. It may be mentioned that \citet{Renardy2000} observed similar time-periodic oscillations for wavenumbers ranging between $5.6$ and $5.8$. Thus, the validation exercise reveals an excellent agreement between the two results, thereby establishing the reliability of our methodology.   

\subsection{Rayleigh-B\'enard convection}

We begin our analysis by considering purely RB convection $(\Lambda_T = 0$, i.e. $Y = 0)$. The critical height ratio, $h^*$, is defined as the height ratio (lower to upper) at which the $Ra$ in both the layers are equal and is given by   

\begin{equation}
h^* = \Bigl({\kappa_r{\kappa_T}_r\eta_r  \over \rho_r \beta_r  }\Bigr)^{({1 \over 4})}, 
\label{EqualRa_criteria}
\end{equation}

\begin{figure}
\centering 
\includegraphics[width=0.45\textwidth]{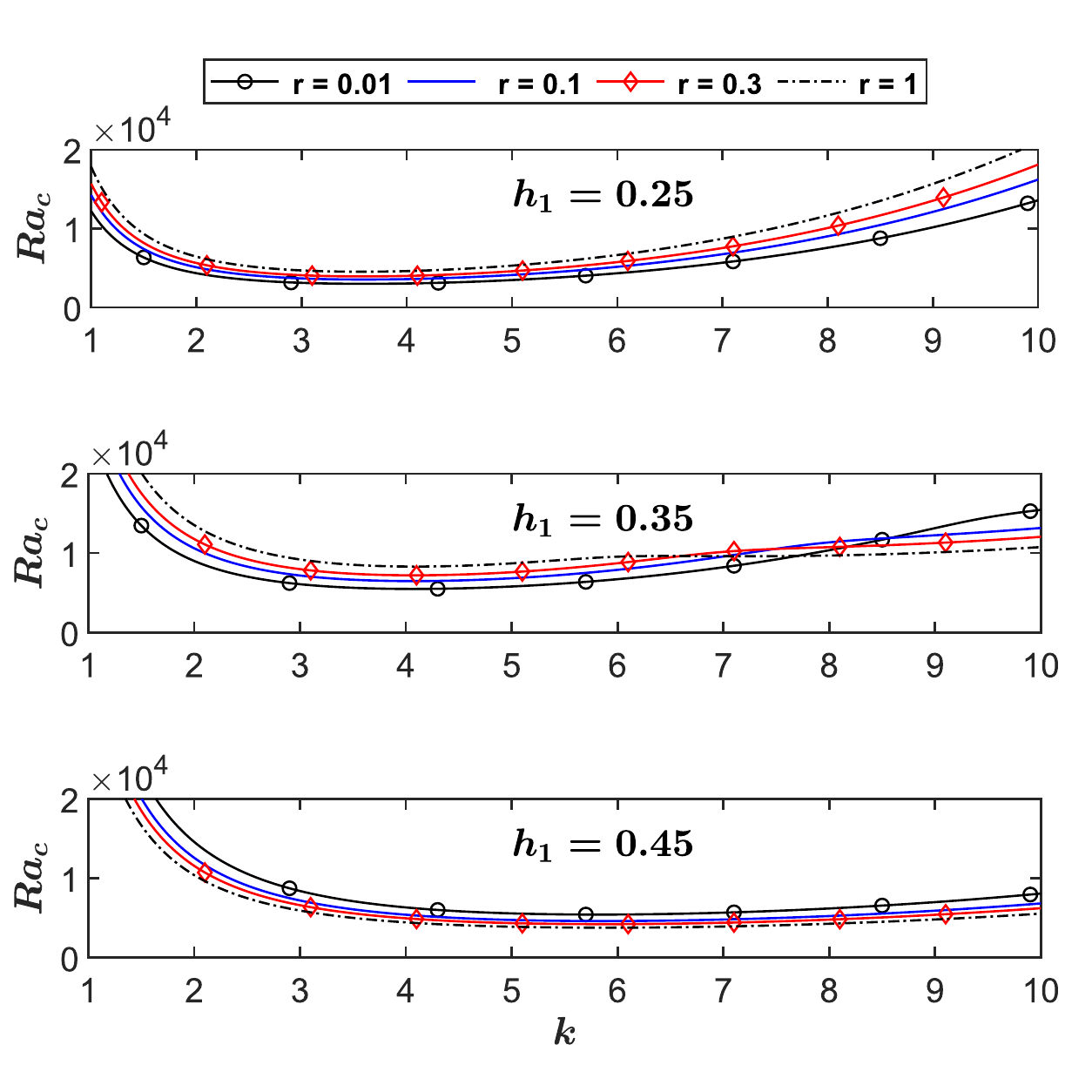} 
\caption{The variation of $Ra_c$ with wavenumber $k$ for the FC72- 1cSt silicone oil system with varying miscibility $(r = 0.01, 0.1, 0.3$ and $1)$ at three different heights of the lower layer $(h_1 = 0.25, 0.35$ and $0.45)$.}
\label{fig8}    
\end{figure}

\begin{figure}
\centering 
\includegraphics[width=0.45\textwidth]{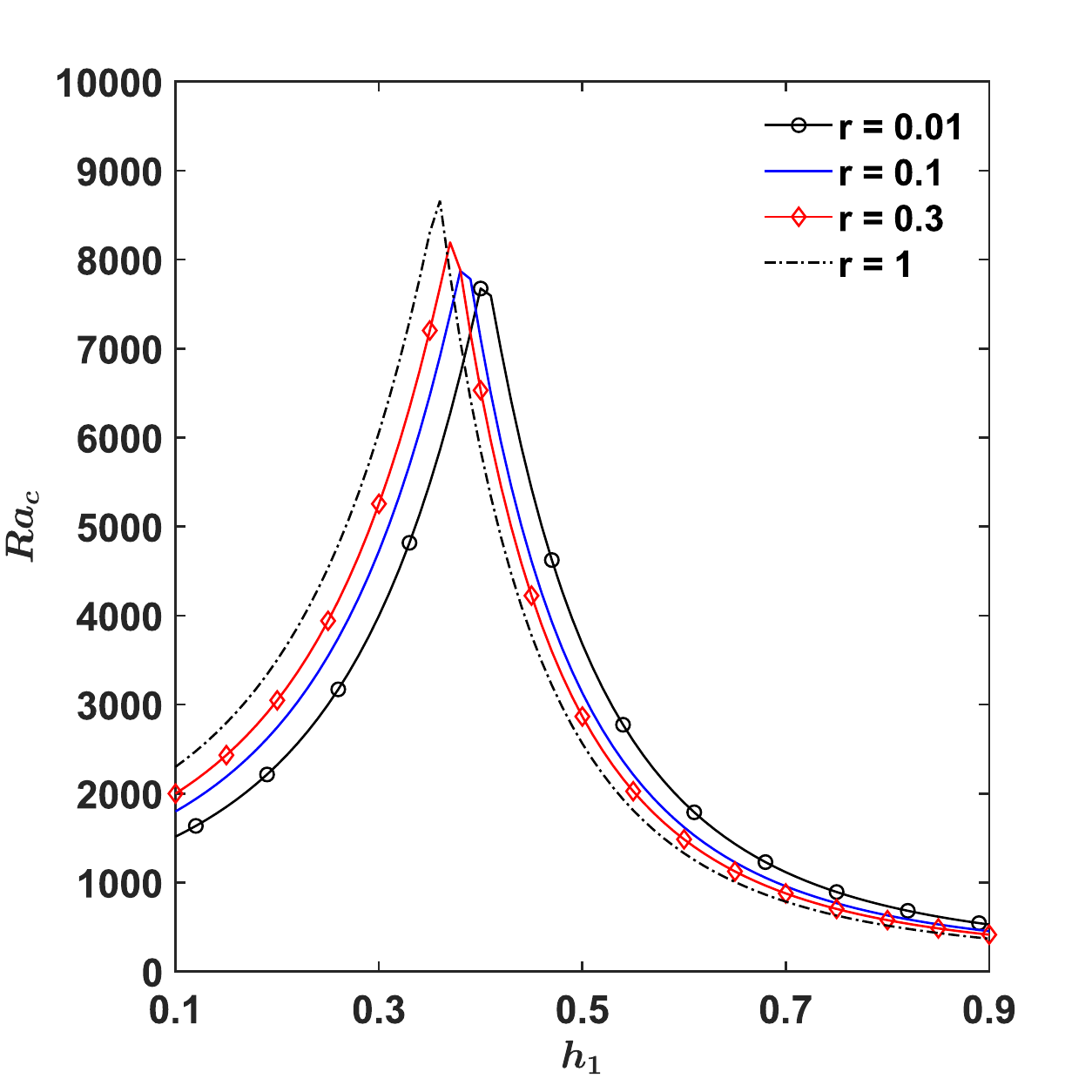} 
\caption{The variation of $Ra_c$ with height of the lower layer $h_1$ for FC72- 1cSt silicone oil with varying miscibility $(r = 0.01, 0.1, 0.3$ and $1)$.}
\label{fig9}    
\end{figure}

\begin{figure}
\centering 
\includegraphics[width=0.45\textwidth]{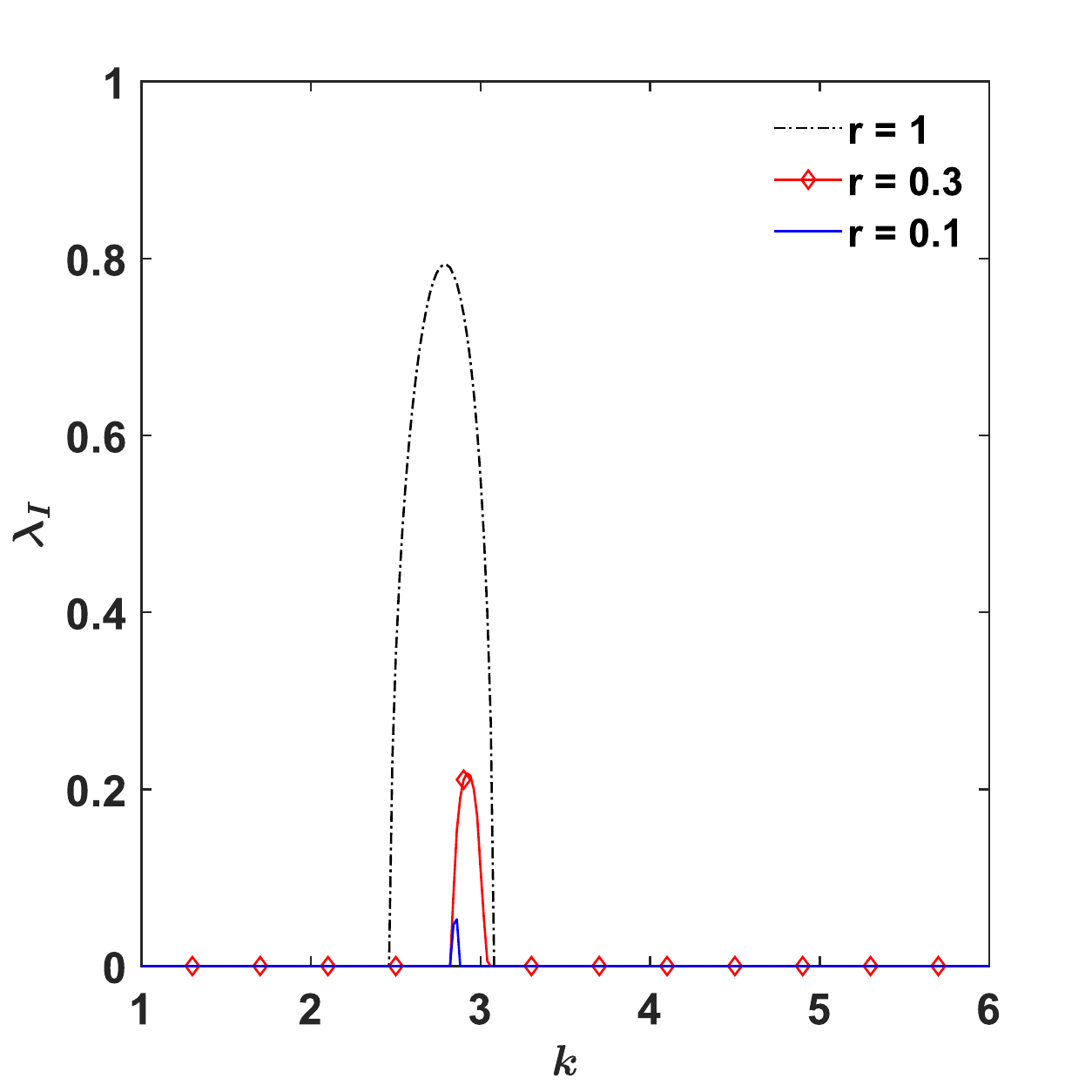} 
\caption{Variation of imaginary component of the growth rate $\lambda_{I}$ for RB convection with miscibility parameter $r$ and height of the lower layer $h_1$ being $(r, h_1)$ = $(0.1,0.41), (0.3,0.40)$ and $(1,0.39)$.}
\label{fig10}    
\end{figure}

\begin{figure*}
\centering 
\includegraphics[width=0.4\textwidth]{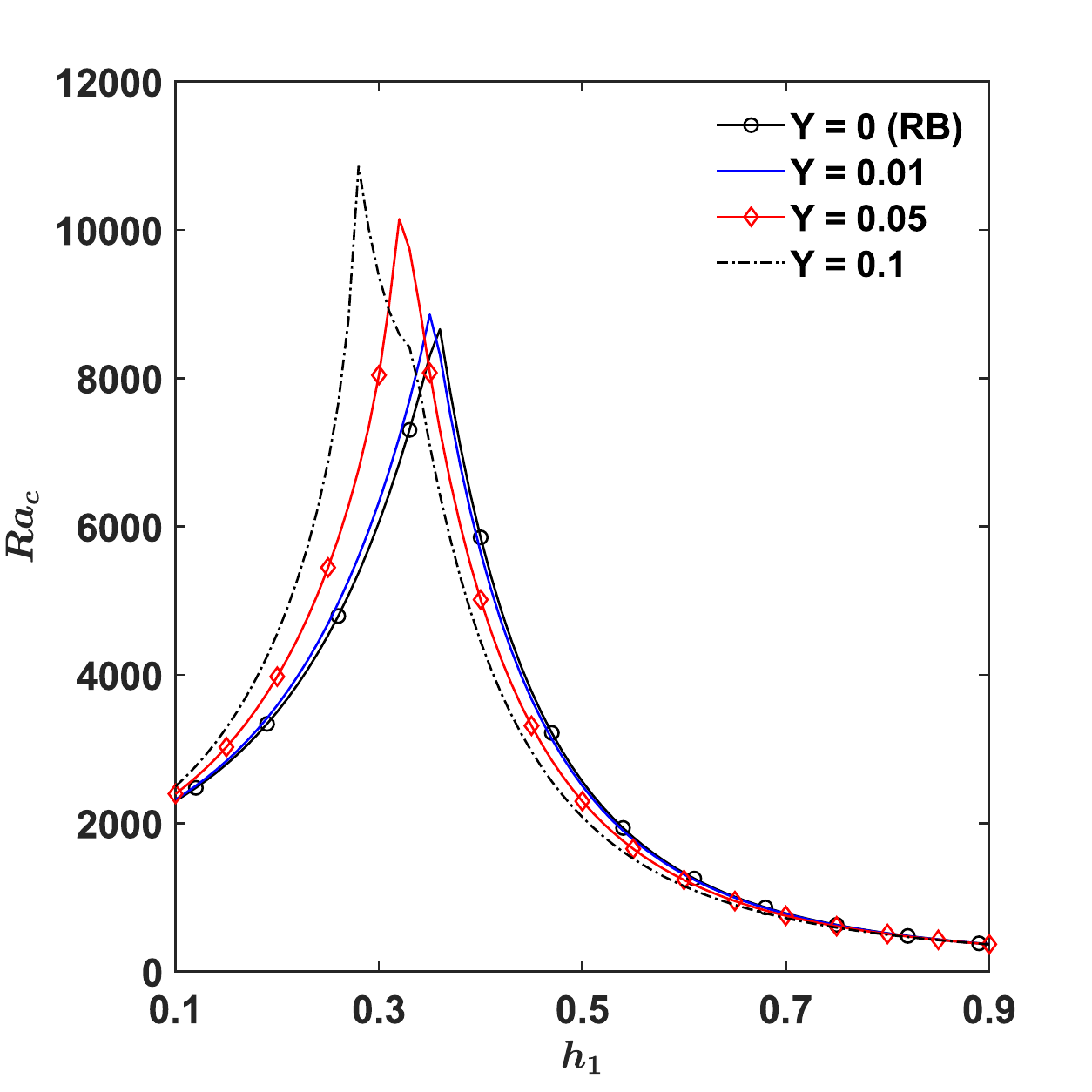} \hspace{0.2cm}
\includegraphics[width=0.4\textwidth]{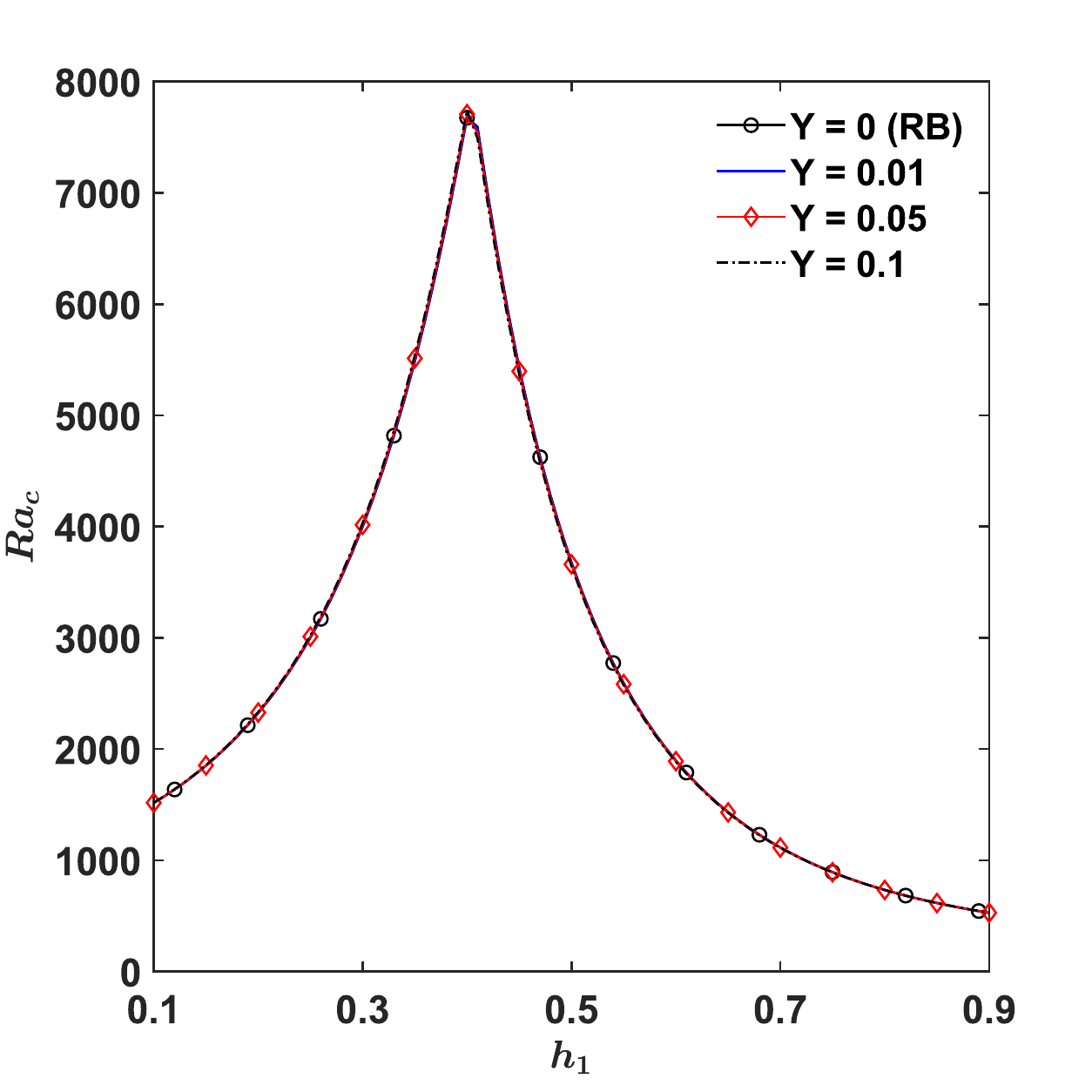} \\
\hspace{0.1cm} (a) \hspace{7.2cm} (b) \\
\caption{Comparison of RB and RBM for $(a) r = 1$ and $(b) r = 0.01$.}
\label{fig11}
\end{figure*}

\begin{figure*}
\centering 
\includegraphics[width=0.4\textwidth]{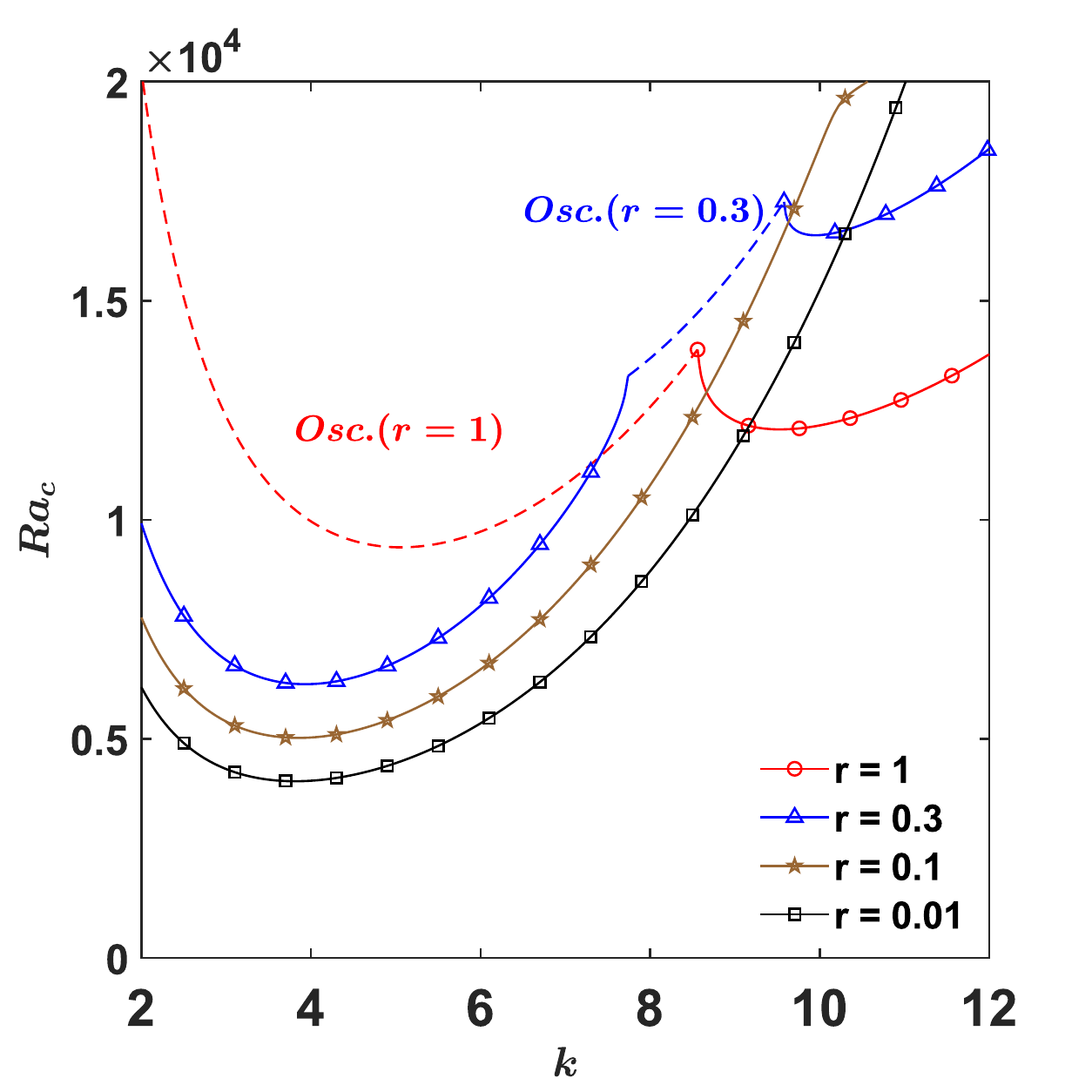} \hspace{0.2cm}
\includegraphics[width=0.4\textwidth]{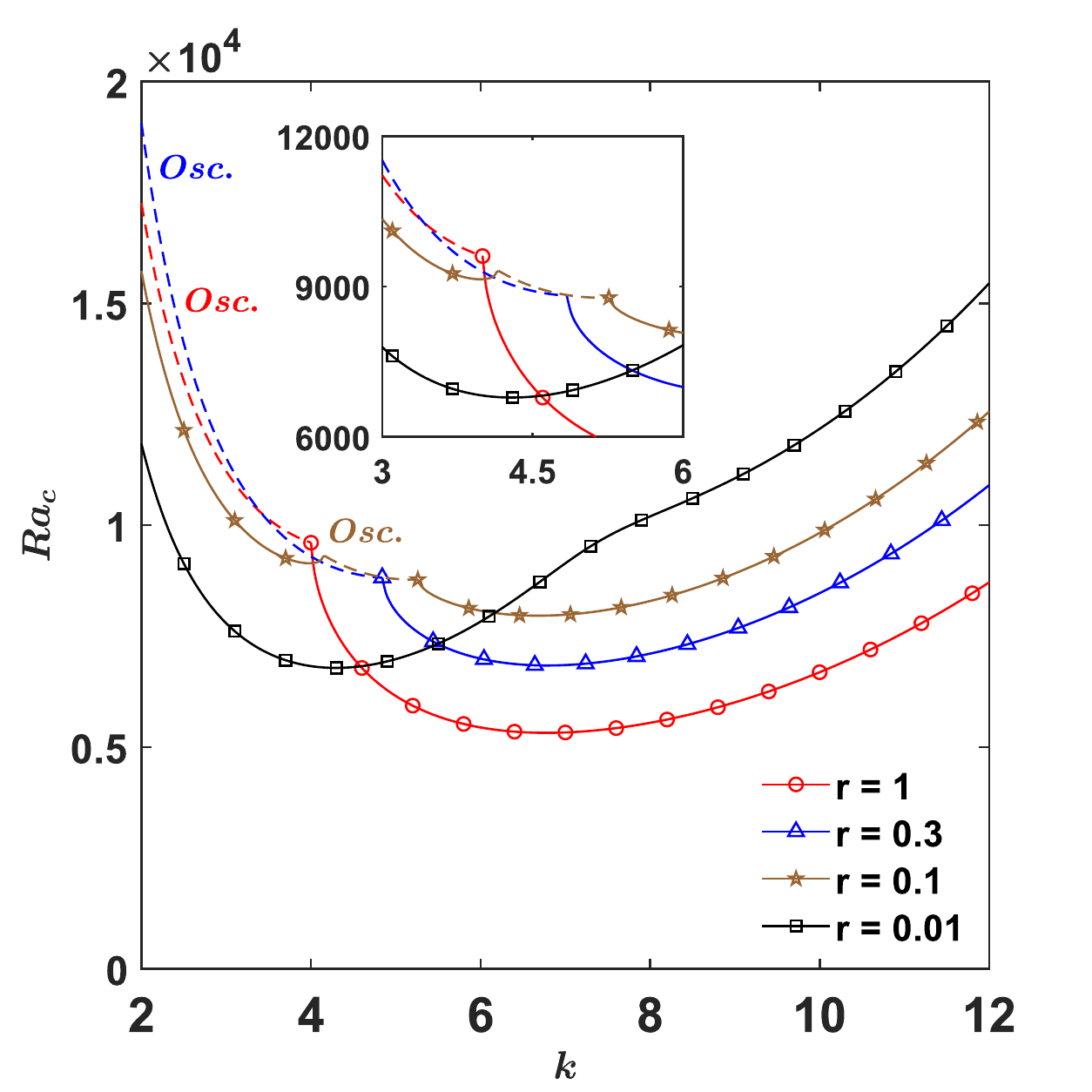} \\
\hspace{0.1cm} (a) \hspace{7.2cm} (b) \\
\caption{The evolution of the window of oscillatory convection for four distinct values of the miscibility parameter $(r = 1, 0.3, 0.1$ and $0.01)$ with $Y = 0.1$ at two different heights of the lower layer $(a) h_1 = 0.3$ and $(b) h_1 = 0.38$. The solid curves correspond to stationary convection regime whereas the dashed segment correspond to oscillatory (Osc.) convection.}
\label{fig12}
\end{figure*}

\begin{figure*}
\centering 
\includegraphics[width=1\textwidth]
{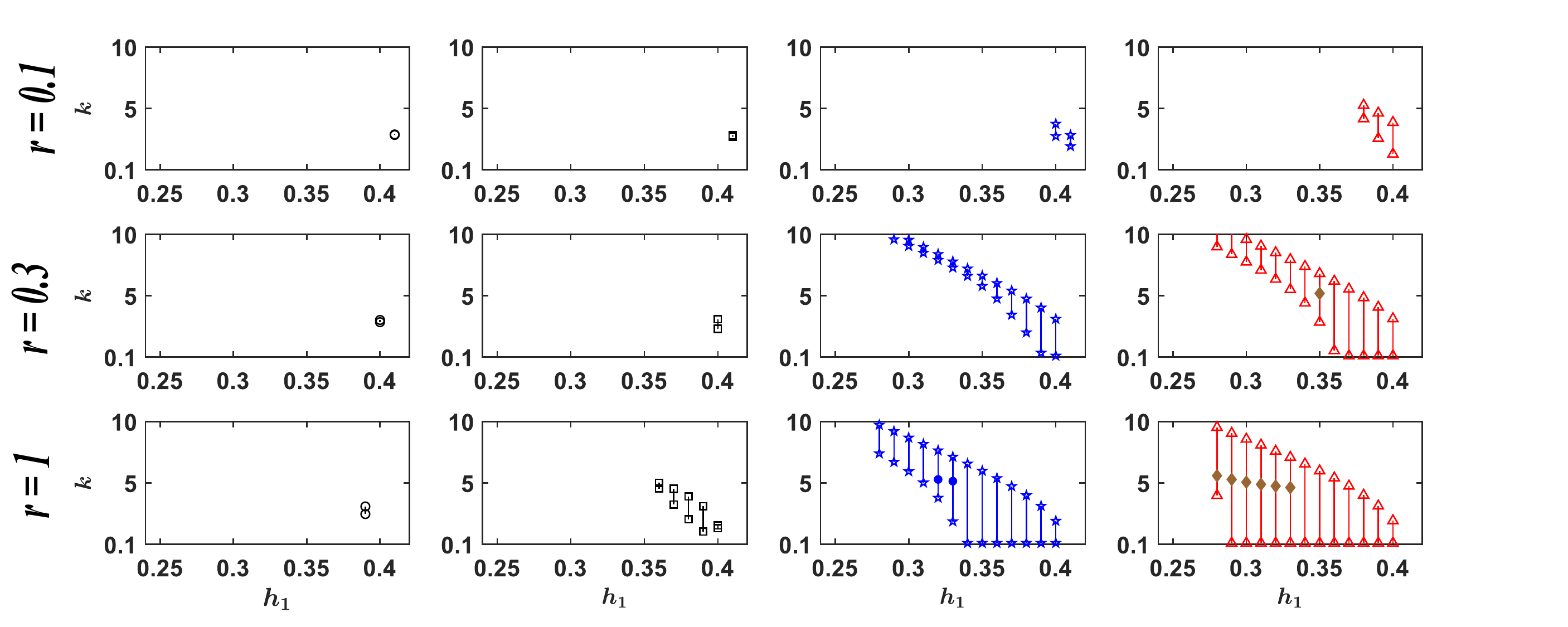} 
\hspace{0cm} (a) \hspace{3.2cm} (b) \hspace{3.2cm} (c) \hspace{3.2cm} (d) \\
\caption{Windows of oscillatory onset depicted as a function of wavenumber $k$, height of the lower layer $h_1$, the miscibility parameter $r$ and the strength of the Marangoni flow $Y$ with $(a)$ $Y = 0(RB)$, $(b)$ $Y = 0.01$, $(c)$ $Y = 0.05$ and $(d)$ $Y = 0.1$. The filled symbols (circles and diamonds) mark the intersection of $k_c$ and the oscillatory window.}
\label{fig13}    
\end{figure*}

For FC72-silicone oil system, $h^* = 0.5532$. For heights resulting in a ratio lower than this $h^*$, i.e., $ h_1 < 0.356$, the upper layer has a higher propensity for convection excitation. Subsequently, the viscous coupling at the interface leads to the lower layer being effectively dragged to satisfy the kinematic condition. The flow is categorized as the so-called upper-dragging mode or the top-dragging mode \cite{Johnson1997}. Thus, in this configuration, the convection onset criteria are primarily governed by the characteristics of the upper layer. On the other hand, if the lower layer height is higher than 0.356, the fluid pair would tend to follow the lower-dragging or bottom-dragging mode. Figure \ref{fig8} depicts the variation of $Ra_c$ with $k$ for three different values of of lower layer height $(h_1 = 0.25, 0.35$ and $0.45)$ at four different values of the miscibility parameter $(r = 0.01, 0.1, 0.3$ and $1)$. For $h_1 = 0.25$, i.e., in the regime of upper-dragging mode, the $Ra_c$ is found to monotonically decrease as the value of the miscibility parameter $r$ is decreased across all the wavenumbers considered. This behaviour can be explained by evaluating the change in effective thermophysical properties of the upper layer (silicone oil). As the system temperature is raised towards the UCST, the degree of miscibility increases, characterized by a decrease in $r$. This increased miscibility would drive the thermophysical properties in the upper (silicone oil-rich) region towards the averaged value of the two fluids. Consequently, the destabilizing agent, i.e. the thermal expansion coefficient, will increase and the stabilizing agents, such as the kinematic viscosity and the thermal diffusivity, will decrease. Thus, the upper region requires a lesser amount of thermal energy for the convective flow onset, thereby decreasing the $Ra_c$. Let us now consider the case of $h_1 = 0.35$, which still corresponds to higher $Ra$ for the upper layer, but is very close to the critical height ratio. Interestingly, we observed a switch in the behaviour of partially miscible cases for relatively short wavelength disturbances compared to the purely immiscible case. An intersection between the $Ra_c$ vs $k$ curves can be seen for $k \in [6.5, 8.5]$. The behaviour at the most dangerous wavenumber ($k$ around $4$) is still the same as that observed for $h_1 = 0.25$. However, the demand for relatively higher thermal energy for short wavelength disturbances at lower values of $r$ highlights the tendency of the system to transit towards the lower-dragging mode. Finally, the $Ra_c$ is investigated for $h_1 = 0.45$. The higher $Ra$ for the lower layer would now provoke lower-dragging mode convection in the bi-layer system, and therefore, the onset criteria are governed by the thermophysical properties of FC72. A complete reversal of the $Ra_c$ vs $k$ variation compared to $h_1 = 0.25$ is observed across all wavenumbers. The evolution of the thermophysical properties of the FC72-rich region as the miscibility is increased leads to enhanced stability of the lower layer. We present the $Ra_c$ as a function of the height of the lower layer in Fig. \ref{fig9}. Evidently, the upper dragging mode is relatively destabilized $($lower $Ra_c)$ as compared to the lower dragging mode owing to the directional evolution of destabilizing and stabilizing agents as a function of partial miscibility. Starting with the asymptotic immiscible configuration $(r = 1$, i.e. far below UCST$)$, we further observe a rightward shift in the maximum value of the $Ra_c$ as the miscibility parameter $r$ is decreased. The combinations of maximum $Ra_c$ and the corresponding height $h_1$ of the lower layer are : $(h_1, Ra_c)$ = $(0.36,8660.88)$, $(0.37,8193.56)$, $(0.38,7883.32)$ and $(0.4,7675.12)$ for $r = 1, 0.3, 0.1$ and $0.01$ respectively. The rightward shift is attributed to the higher thickness of the lower layer required to have similar $Ra$ in both layers.\\

The onset of the convection is non-oscillatory in nature, irrespective of the height of the lower layer. However, for certain values of $h_1$, we found a window of wavenumbers, excluding the most dangerous wavenumber, to be unstable to oscillatory onset. The time period $(\tau)$ of oscillation can be calculated using the expression \cite{RenardyYY1996}:      

\begin{equation}
\tau = \Bigl({2 \pi {H_2}^2  \over {\kappa_T}_2 | \lambda_{I} | }\Bigr), 
\label{timePeriod}
\end{equation}

Figure \ref{fig10} depicts the variation of the imaginary component of the growth rate with the wavenumber. As the miscibility parameter $r$ decreases, the window for plausible oscillatory onset shrinks. The time period $\tau$ of oscillation is found to be approximately $44$, $145$ and $612$ minutes for $(r,h_1) = (1,0.39), (0.3,0.40)$ and $(0.1,0.41)$ respectively. The oscillations occur due to a switch in the convection modes between mechanical coupling mode and thermal convection mode. The co-rotating rolls in thermal coupling mode transform into the counter-rotating rolls in mechanical coupling mode via the formation of an eddy roll between the two fluids. The increased time period of oscillation with an increasing degree of miscibility is attributed to the highly diffuse nature of the interface, which impedes the formation of these eddy rolls. We did not observe any oscillations for $r = 0.01$. The increase in the degree of partial miscibility leads to the reduction of the ratio of thermophysical properties. Thus, the system is effectively driven away from the prospects of oscillations as suggested by the ratio combination criteria \cite{RenardyYY1996}. Further, the increase in the thermal diffusivity of the lower layer with an increase in miscibility impedes the heat transfer from the lower boundary to the interfacial region. Having developed an understanding of the RB convection in the FC72 - silicone oil system, we present our findings for the RBM convection in the ensuing section.        

\subsection{Rayleigh-B\'enard-Marangoni convection}

We now perform a parametric analysis of the RBM convection by considering four distinct values of $Y = (0, 0.01, 0.05$ and $0.1)$ highlighting the strength of the surface tension gradient component over the buoyancy force; $Y = 0$ signifies the absence of thermo-capillary flow therefore pure RB convection. Analogously, four distinct degrees of partial miscibility characterized by $r = 1, 0.3, 0.1$ and $0.01$ are considered. We compare RB convection with RBM convection in Fig. \ref{fig11}. The thermo-capillary flow plays a stabilizing role when the interface is situated close to the bottom plate, i.e., during the upper dragging mode, whereas it plays a destabilizing role during the lower dragging mode, as shown in Fig. \ref{fig11}(a) corroborating the findings of \citet{Neopmnyashchy2004} for immiscible liquids $(r = 1)$. During the lower dragging mode, the convection is initiated in the lower layer owing to the buoyancy forces. This convective flow, in turn, establishes a temperature gradient along the interface as the hot rising fluid in the bottom layer creates a localized hot spot at the interface. The surface tension between the FC72-silicone oil decreases linearly with temperature, thereby creating a gradient of surface tension between the hot spots and the relatively cold spots. Thus, a Marangoni flow emanates, which supports the primary buoyant flow in the lower layer (see also fig. \ref{fig1}). Consequently, the higher the strength of this Marangoni flow, the lower the energy required to provoke the system into convection, hence the destabilizing effect. On the other hand, during the upper dragging mode, the tangential stress arising due to the surface tension gradient tends to counteract the buoyant flow, thus acting as a stabilizing agent. However, when the system is considered at a temperature close to the UCST $(r = 0.01)$, the concentration gradient in the interfacial region is very low. Therefore, the Korteweg stresses mimicking the effects of surface tension forces are feeble. Consequently, the gradient of surface tension along the interface is also feeble. Fig. \ref{fig11}(b) demonstrates the independence of $Ra_c$ to the Marangoni flow for a bi-layer system considered close to the UCST. The inclusion of the thermo-capillary flow further plays a crucial role in governing the window of oscillatory onset of the convection. Interestingly, for certain height ratios, the onset is found to be oscillatory for $r \ge 0.1$ for all the values of $Y$ considered. However, the onset is stationary for $r = 0.01$, which is expected as the surface tension effects would be minimal with diminishing $r$. For the current FC72 - silicone oil system, the window of height fractions corresponding to oscillatory onset monotonically increases with increasing $Y$. When the fluids are considered at the immiscible limit, $r = 1$, we get oscillatory onset for the combinations $(Y, h_1)$ =  $(0.05, [0.32,0.33])$ and $(0.1,[0.28,0.33])$. As the strength of the Marangoni flow increases, the height fraction values that are unstable to oscillatory onset decrease. \\

Figure \ref{fig12} depicts the evolution of the window of wavenumbers for $Y = 0.1$ that leads to imaginary eigenvalues, thereby signifying the oscillatory convection. The subsequent increase in the degree of partial miscibility monotonically shrinks this window irrespective of the height of the lower layer. The spatio-temporal growth of the perturbation imparted to the system is naturally a function of the wavenumber. For $h_1 = 0.3$, as shown in Fig. \ref{fig12}(a), the plausibility of oscillation vanishes as the degree of partial miscibility is increased beyond a threshold limit as depicted by $r = 0.1$ and $r = 0.01$. On the other hand, a system considered relatively far from the UCST exhibits an interesting wavenumber dependence. At the immiscible limit $(r = 1)$, the most-dangerous wavenumber $(k_c = 5.06)$ falls within the window of oscillatory convection $k = [0.10,8.56]$. Thus, a random disturbance to the binary fluid system would lead to the growth of the amplitude of the perturbation, which corresponds to $k_c$, given that enough thermal energy characterized by $Ra_c$ is provided. Consequently, an oscillatory flow pattern would emerge. However, if the same pair of fluids undergo a certain degree of partial miscibility $(r = 0.3)$, the window of wavenumbers responsible for oscillatory convection ranges between $k = [7.74,9.58]$. This observation further corroborates the role of surface tension in the emanation of oscillatory convection. The increase in the degree of miscibility is marked by a decrease in the surface tension. Thus, a relatively higher wavenumber is required to achieve the interfacial curvature for the enactment of adequate surface tension force. For $r = 0.3$, the most dangerous wavenumber is $k_c = 3.92$, and therefore, a random disturbance would result in the onset of stationary convection. Nevertheless, the imposition of a controlled disturbance such that the wavenumber of the perturbation falls in the range $k = [7.74,9.58]$ would lead to the onset of oscillatory convection. Similarly, Fig. \ref{fig12}(b) depicts the $Ra_c$ vs $k$ variation for $h_1 = 0.38$. Notably, no intersection of the $k_c$ and the oscillation window is observed. Thus, the height of the lower layer or the depth fraction of the two layers plays a key role in determining the propensity of a binary fluid system perturbed with random disturbance towards the oscillatory onset of the convection. \\
We encapsulate the overall picture as a function of wavenumber, height of the lower layer, the degree of partial miscibility and $Y$ in Fig. \ref{fig13}. The unfilled symbols depict the range of wavenumbers that corresponds to oscillatory convection. The filled symbols are the intersection points between the $k_c$ and the above-mentioned range. The solid lines depict the range of wavenumbers that correspond to oscillatory convection. The filled symbols,  circles in the case $r = 1; Y = 0.05$ and diamonds in the case $r = 1; Y = 0.1$ demonstrate that the most dangerous wavenumber $k_c$ falls in the previously obtained oscillation-susceptible range of wavenumbers. Thus, any random perturbation would lead to emanation of oscillatory convection. For all values of the height of the lower layer, we observed a consistent shrink in the propensity of oscillatory onset as the binary fluid system approaches the UCST. For $r = 0.3$, an open-ended window of imaginary eigenvalues is found as the wavenumber increases, for relatively higher values of $Y (= 0.1)$ as shown by red lines for $h_1 = 0.27, 0.28$ and $0.29$. The present work is limited to a maximum wavenumber of $k = 12$. The further increase in the degree of miscibility characterized by $r = 0.1$ leads to a subsequent decrease in the window of plausible oscillatory onset. For $r = 0.01$ (not shown schematically), we did not find any wavenumber that shall lead to oscillation.

\section{Conclusion}
\label{sec:conclusion}

This study analyzes the effect of temperature-dependent solubility in a two-layer system containing FC72- 1cSt silicone oil on the modes of RBM convection. A modified phase-field-based model is employed to track the evolution of the diffuse interface and the equilibrium configuration of the two fluids. The model accurately represents the behaviour of the FC72- 1cSt silicone oil system in the vicinity of UCST. A rigorous comparison with the stability results of \citet{Renardy2000} establishes the ability of the present model in the asymptotic immiscible limit. Here, we have limited our analysis to system temperatures that are lower than the UCST to preserve the definition of the interface and the bi-layer setup. The lower critical temperature differences observed experimentally\cite{Degen} between the bottom and top plates have allowed us to presently assume a constant value for the miscibility parameter $r$, thus significantly reducing the associated mathematical complexities. However, this assumption is not valid for systems extremely close to the UCST. \\
A pseudo-spectral collocation-based method is deployed to solve the linearized equations and obtain the onset characteristics of the convection. The $Ra_c$ is found to be strongly dependent on the miscibility parameter $r$. The directional evolution of the thermophysical properties of the bulk phases with the variation in $r$ governs the relative change in $Ra_c$. In the case of pure buoyancy-driven convection, the divergence of the interfacial thickness as the system approaches the UCST impedes the formation of the eddy roll, which is responsible for switching the convection pattern from thermal coupling mode to mechanical coupling mode. This resulted in a monotonic rise in the time period of oscillation, eventually leading to stationary convection close to the UCST. In the case where the additional effects of thermo-capillarity are considered, the window of oscillatory convection was found to increase. However, as the degree of miscibility between the two fluids increases, the window of wavenumber entailing oscillations decreases. This study, therefore, reveals the particular combination of the pertinent parameters, namely, system temperature and the relative strength of the Marangoni flow required to sample the oscillatory onset of convection. Thus, control over the phenomenon of micro-droplet formation \cite{Aibara2020} can be achieved if the system is perturbed with an apriori known wavenumber.

\begin{acknowledgments}
We thank the organization of the International Research Network (IRN) under the auspice of CNRS for the workshop held in IIT Madras, India on 25-27 October 2024. This IRN has brought together the audience to which, this work was first presented.
\end{acknowledgments}

\section*{Funding}
A.D. and S.A. gratefully acknowledge the support provided by CNES (Centre national d'études spatiales: grant 9384- 4500082826). 

\section*{Conflict of Interest}
The authors have no conflict of interest to disclose.

\section*{Data Availability}
The data that support the findings of this study are available from the corresponding author upon request.

\appendix

\nocite{*}
\bibliography{main}

\begin{thebibliography}{40}%
\makeatletter
\providecommand \@ifxundefined [1]{%
 \@ifx{#1\undefined}
}%
\providecommand \@ifnum [1]{%
 \ifnum #1\expandafter \@firstoftwo
 \else \expandafter \@secondoftwo
 \fi
}%
\providecommand \@ifx [1]{%
 \ifx #1\expandafter \@firstoftwo
 \else \expandafter \@secondoftwo
 \fi
}%
\providecommand \natexlab [1]{#1}%
\providecommand \enquote  [1]{``#1''}%
\providecommand \bibnamefont  [1]{#1}%
\providecommand \bibfnamefont [1]{#1}%
\providecommand \citenamefont [1]{#1}%
\providecommand \href@noop [0]{\@secondoftwo}%
\providecommand \href [0]{\begingroup \@sanitize@url \@href}%
\providecommand \@href[1]{\@@startlink{#1}\@@href}%
\providecommand \@@href[1]{\endgroup#1\@@endlink}%
\providecommand \@sanitize@url [0]{\catcode `\\12\catcode `\$12\catcode `\&12\catcode `\#12\catcode `\^12\catcode `\_12\catcode `\%12\relax}%
\providecommand \@@startlink[1]{}%
\providecommand \@@endlink[0]{}%
\providecommand \url  [0]{\begingroup\@sanitize@url \@url }%
\providecommand \@url [1]{\endgroup\@href {#1}{\urlprefix }}%
\providecommand \urlprefix  [0]{URL }%
\providecommand \Eprint [0]{\href }%
\providecommand \doibase [0]{http://dx.doi.org/}%
\providecommand \selectlanguage [0]{\@gobble}%
\providecommand \bibinfo  [0]{\@secondoftwo}%
\providecommand \bibfield  [0]{\@secondoftwo}%
\providecommand \translation [1]{[#1]}%
\providecommand \BibitemOpen [0]{}%
\providecommand \bibitemStop [0]{}%
\providecommand \bibitemNoStop [0]{.\EOS\space}%
\providecommand \EOS [0]{\spacefactor3000\relax}%
\providecommand \BibitemShut  [1]{\csname bibitem#1\endcsname}%
\let\auto@bib@innerbib\@empty
\bibitem [{\citenamefont {Schmaljohann}(2006)}]{Schmaljohann2006}%
  \BibitemOpen
  \bibfield  {author} {\bibinfo {author} {\bibfnamefont {D.}~\bibnamefont {Schmaljohann}},\ }\bibfield  {title} {\enquote {\bibinfo {title} {Thermo- and ph-responsive polymers in drug delivery},}\ }\href@noop {} {\bibfield  {journal} {\bibinfo  {journal} {Adv.\ Drug Deliv.\ Rev.}\ }\textbf {\bibinfo {volume} {58}},\ \bibinfo {pages} {1655--1670} (\bibinfo {year} {2006})}\BibitemShut {NoStop}%
\bibitem [{\citenamefont {Kohno}\ \emph {et~al.}(2011)\citenamefont {Kohno}, \citenamefont {Saita}, \citenamefont {Murata}, \citenamefont {Nakamura},\ and\ \citenamefont {Ohno}}]{Kohno2011}%
  \BibitemOpen
  \bibfield  {author} {\bibinfo {author} {\bibfnamefont {Y.}~\bibnamefont {Kohno}}, \bibinfo {author} {\bibfnamefont {S.}~\bibnamefont {Saita}}, \bibinfo {author} {\bibfnamefont {K.}~\bibnamefont {Murata}}, \bibinfo {author} {\bibfnamefont {N.}~\bibnamefont {Nakamura}}, \ and\ \bibinfo {author} {\bibfnamefont {H.}~\bibnamefont {Ohno}},\ }\bibfield  {title} {\enquote {\bibinfo {title} {Extraction of proteins with temperature sensitive and reversible phase change of ionic liquid/water mixture},}\ }\href@noop {} {\bibfield  {journal} {\bibinfo  {journal} {Polym.\ Chem.}\ }\textbf {\bibinfo {volume} {2}},\ \bibinfo {pages} {862--867} (\bibinfo {year} {2011})}\BibitemShut {NoStop}%
\bibitem [{\citenamefont {Ventura}\ \emph {et~al.}(2017)\citenamefont {Ventura}, \citenamefont {Silva}, \citenamefont {Quental}, \citenamefont {Freire},\ and\ \citenamefont {Coutinho}}]{Ventura2017}%
  \BibitemOpen
  \bibfield  {author} {\bibinfo {author} {\bibfnamefont {S.}~\bibnamefont {Ventura}}, \bibinfo {author} {\bibfnamefont {F.}~\bibnamefont {Silva}}, \bibinfo {author} {\bibfnamefont {M.}~\bibnamefont {Quental}}, \bibinfo {author} {\bibfnamefont {D.~M.~M.}\ \bibnamefont {Freire}}, \ and\ \bibinfo {author} {\bibfnamefont {J.}~\bibnamefont {Coutinho}},\ }\bibfield  {title} {\enquote {\bibinfo {title} {Ionic-liquid-mediated extraction and separation processes for bioactive compounds: Past, present and future trends},}\ }\href@noop {} {\bibfield  {journal} {\bibinfo  {journal} {Chem. Rev.}\ }\textbf {\bibinfo {volume} {117}},\ \bibinfo {pages} {6984--7052} (\bibinfo {year} {2017})}\BibitemShut {NoStop}%
\bibitem [{\citenamefont {Dessimoz}\ \emph {et~al.}(2008)\citenamefont {Dessimoz}, \citenamefont {Cavin}, \citenamefont {Renken},\ and\ \citenamefont {Kiwi-Minsker}}]{Dessimoz2008}%
  \BibitemOpen
  \bibfield  {author} {\bibinfo {author} {\bibfnamefont {A.}~\bibnamefont {Dessimoz}}, \bibinfo {author} {\bibfnamefont {L.}~\bibnamefont {Cavin}}, \bibinfo {author} {\bibfnamefont {A.}~\bibnamefont {Renken}}, \ and\ \bibinfo {author} {\bibfnamefont {L.}~\bibnamefont {Kiwi-Minsker}},\ }\bibfield  {title} {\enquote {\bibinfo {title} {Liquid-liquid two phase flow patterns and mass transfer characteristics in rectangular glass microreactors},}\ }\href@noop {} {\bibfield  {journal} {\bibinfo  {journal} {Chem.\ Eng.\ Sci.}\ }\textbf {\bibinfo {volume} {63}},\ \bibinfo {pages} {4035--4044} (\bibinfo {year} {2008})}\BibitemShut {NoStop}%
\bibitem [{\citenamefont {Patel}\ \emph {et~al.}(2021)\citenamefont {Patel}, \citenamefont {Radhakrishnan}, \citenamefont {Bescher}, \citenamefont {Hunter-Sellars}, \citenamefont {Schmidt-Hansberg}, \citenamefont {Amstad}, \citenamefont {Ibsen},\ and\ \citenamefont {Guldin}}]{Patel2021}%
  \BibitemOpen
  \bibfield  {author} {\bibinfo {author} {\bibfnamefont {M.}~\bibnamefont {Patel}}, \bibinfo {author} {\bibfnamefont {A.}~\bibnamefont {Radhakrishnan}}, \bibinfo {author} {\bibfnamefont {L.}~\bibnamefont {Bescher}}, \bibinfo {author} {\bibfnamefont {E.}~\bibnamefont {Hunter-Sellars}}, \bibinfo {author} {\bibfnamefont {B.}~\bibnamefont {Schmidt-Hansberg}}, \bibinfo {author} {\bibfnamefont {E.}~\bibnamefont {Amstad}}, \bibinfo {author} {\bibfnamefont {S.}~\bibnamefont {Ibsen}}, \ and\ \bibinfo {author} {\bibfnamefont {S.}~\bibnamefont {Guldin}},\ }\bibfield  {title} {\enquote {\bibinfo {title} {Temperature-induced liquid crystal microdroplet formation in a partially miscible liquid mixture},}\ }\href@noop {} {\bibfield  {journal} {\bibinfo  {journal} {Soft Matter}\ }\textbf {\bibinfo {volume} {17}},\ \bibinfo {pages} {947} (\bibinfo {year} {2021})}\BibitemShut {NoStop}%
\bibitem [{\citenamefont {Joseph}(1990)}]{Joseph1990}%
  \BibitemOpen
  \bibfield  {author} {\bibinfo {author} {\bibfnamefont {D.}~\bibnamefont {Joseph}},\ }\bibfield  {title} {\enquote {\bibinfo {title} {Fluid dynamics of two miscible liquids with slow diffusion and korteweg stresses},}\ }\href@noop {} {\bibfield  {journal} {\bibinfo  {journal} {Eur.\ J.\ Mech.\ B/Fluids}\ }\textbf {\bibinfo {volume} {9(6)}},\ \bibinfo {pages} {565--596} (\bibinfo {year} {1990})}\BibitemShut {NoStop}%
\bibitem [{\citenamefont {Fornerod}, \citenamefont {Amstad},\ and\ \citenamefont {Guldin}(2020)}]{Fornerod2020}%
  \BibitemOpen
  \bibfield  {author} {\bibinfo {author} {\bibfnamefont {M.}~\bibnamefont {Fornerod}}, \bibinfo {author} {\bibfnamefont {E.}~\bibnamefont {Amstad}}, \ and\ \bibinfo {author} {\bibfnamefont {S.}~\bibnamefont {Guldin}},\ }\bibfield  {title} {\enquote {\bibinfo {title} {Microfluidics of binary liquid mixtures with temperature-dependent miscibility},}\ }\href@noop {} {\bibfield  {journal} {\bibinfo  {journal} {Mol.\ Syst.\ Des.\ Eng.}\ }\textbf {\bibinfo {volume} {5}},\ \bibinfo {pages} {358} (\bibinfo {year} {2020})}\BibitemShut {NoStop}%
\bibitem [{\citenamefont {Aibara}\ \emph {et~al.}(2020)\citenamefont {Aibara}, \citenamefont {Katoh}, \citenamefont {Minamoto}, \citenamefont {Uwada},\ and\ \citenamefont {Hashimoto}}]{Aibara2020}%
  \BibitemOpen
  \bibfield  {author} {\bibinfo {author} {\bibfnamefont {I.}~\bibnamefont {Aibara}}, \bibinfo {author} {\bibfnamefont {T.}~\bibnamefont {Katoh}}, \bibinfo {author} {\bibfnamefont {C.}~\bibnamefont {Minamoto}}, \bibinfo {author} {\bibfnamefont {T.}~\bibnamefont {Uwada}}, \ and\ \bibinfo {author} {\bibfnamefont {S.}~\bibnamefont {Hashimoto}},\ }\bibfield  {title} {\enquote {\bibinfo {title} {Liquid-liquid interface can promote micro-scale thermal marangoni convection in liquid binary mixtures},}\ }\href@noop {} {\bibfield  {journal} {\bibinfo  {journal} {J.\ Phys.\ Chem.\ C}\ }\textbf {\bibinfo {volume} {124}},\ \bibinfo {pages} {2427--2438} (\bibinfo {year} {2020})}\BibitemShut {NoStop}%
\bibitem [{\citenamefont {Johnson}(1975)}]{Johnson1975}%
  \BibitemOpen
  \bibfield  {author} {\bibinfo {author} {\bibfnamefont {E.}~\bibnamefont {Johnson}},\ }\bibfield  {title} {\enquote {\bibinfo {title} {Liquid encapsulated floating zone melting of gaas},}\ }\href@noop {} {\bibfield  {journal} {\bibinfo  {journal} {J. Cryst.\ Growth}\ }\textbf {\bibinfo {volume} {30}},\ \bibinfo {pages} {249--256} (\bibinfo {year} {1975})}\BibitemShut {NoStop}%
\bibitem [{\citenamefont {Busse}(1981)}]{Busse1981}%
  \BibitemOpen
  \bibfield  {author} {\bibinfo {author} {\bibfnamefont {F.}~\bibnamefont {Busse}},\ }\bibfield  {title} {\enquote {\bibinfo {title} {On the aspect ratios of two layer mantle convection},}\ }\href@noop {} {\bibfield  {journal} {\bibinfo  {journal} {Phys.\ Earth Planet.\ Inter.}\ }\textbf {\bibinfo {volume} {24}},\ \bibinfo {pages} {320--324} (\bibinfo {year} {1981})}\BibitemShut {NoStop}%
\bibitem [{\citenamefont {Zeren}\ and\ \citenamefont {Reynolds}(1964)}]{Zeren1972}%
  \BibitemOpen
  \bibfield  {author} {\bibinfo {author} {\bibfnamefont {R.}~\bibnamefont {Zeren}}\ and\ \bibinfo {author} {\bibfnamefont {W.}~\bibnamefont {Reynolds}},\ }\bibfield  {title} {\enquote {\bibinfo {title} {Thermal instabilities in two-fluid horizontal layers},}\ }\href@noop {} {\bibfield  {journal} {\bibinfo  {journal} {J. Fluid Mech.}\ }\textbf {\bibinfo {volume} {53}},\ \bibinfo {pages} {305--327} (\bibinfo {year} {1964})}\BibitemShut {NoStop}%
\bibitem [{\citenamefont {Johnson}\ and\ \citenamefont {Narayanan}(1997)}]{Johnson1997}%
  \BibitemOpen
  \bibfield  {author} {\bibinfo {author} {\bibfnamefont {D.}~\bibnamefont {Johnson}}\ and\ \bibinfo {author} {\bibfnamefont {R.}~\bibnamefont {Narayanan}},\ }\bibfield  {title} {\enquote {\bibinfo {title} {Geometric effects on convective coupling and interfacial structures in bilayer convection},}\ }\href@noop {} {\bibfield  {journal} {\bibinfo  {journal} {Phys.\ Rev.\ E}\ }\textbf {\bibinfo {volume} {56(5)}},\ \bibinfo {pages} {5462(11)} (\bibinfo {year} {1997})}\BibitemShut {NoStop}%
\bibitem [{\citenamefont {Rasenat}, \citenamefont {Busse},\ and\ \citenamefont {Rehberg}(1989)}]{Rasenat1989}%
  \BibitemOpen
  \bibfield  {author} {\bibinfo {author} {\bibfnamefont {S.}~\bibnamefont {Rasenat}}, \bibinfo {author} {\bibfnamefont {F.}~\bibnamefont {Busse}}, \ and\ \bibinfo {author} {\bibfnamefont {I.}~\bibnamefont {Rehberg}},\ }\bibfield  {title} {\enquote {\bibinfo {title} {A theoretical and experimental study of double-layer convection},}\ }\href@noop {} {\bibfield  {journal} {\bibinfo  {journal} {J. Fluid Mech.}\ }\textbf {\bibinfo {volume} {199}},\ \bibinfo {pages} {519--540} (\bibinfo {year} {1989})}\BibitemShut {NoStop}%
\bibitem [{\citenamefont {Renardy}\ and\ \citenamefont {Joseph}(1985)}]{RenardyJoseph1985}%
  \BibitemOpen
  \bibfield  {author} {\bibinfo {author} {\bibfnamefont {Y.}~\bibnamefont {Renardy}}\ and\ \bibinfo {author} {\bibfnamefont {D.}~\bibnamefont {Joseph}},\ }\bibfield  {title} {\enquote {\bibinfo {title} {Oscillatory instability in a b\'enard problem of two fluids},}\ }\href@noop {} {\bibfield  {journal} {\bibinfo  {journal} {Phys. Fluids}\ }\textbf {\bibinfo {volume} {28}},\ \bibinfo {pages} {788--793} (\bibinfo {year} {1985})}\BibitemShut {NoStop}%
\bibitem [{\citenamefont {Renardy}\ and\ \citenamefont {Renardy}(1985)}]{Renardy1985}%
  \BibitemOpen
  \bibfield  {author} {\bibinfo {author} {\bibfnamefont {Y.}~\bibnamefont {Renardy}}\ and\ \bibinfo {author} {\bibfnamefont {M.}~\bibnamefont {Renardy}},\ }\bibfield  {title} {\enquote {\bibinfo {title} {Perturbation analysis of steady and oscillatory onset in a b\'enard problem with two similar liquids},}\ }\href@noop {} {\bibfield  {journal} {\bibinfo  {journal} {Phys. Fluids}\ }\textbf {\bibinfo {volume} {28}},\ \bibinfo {pages} {2699--2708} (\bibinfo {year} {1985})}\BibitemShut {NoStop}%
\bibitem [{\citenamefont {Colinet}\ and\ \citenamefont {Legros}(1994)}]{Colinet1994}%
  \BibitemOpen
  \bibfield  {author} {\bibinfo {author} {\bibfnamefont {P.}~\bibnamefont {Colinet}}\ and\ \bibinfo {author} {\bibfnamefont {J.}~\bibnamefont {Legros}},\ }\bibfield  {title} {\enquote {\bibinfo {title} {On the hopf bifurcation occuring in the two layer rayleigh-b\'enard convective instability},}\ }\href@noop {} {\bibfield  {journal} {\bibinfo  {journal} {Phys. Fluids}\ }\textbf {\bibinfo {volume} {6}},\ \bibinfo {pages} {2631--2639} (\bibinfo {year} {1994})}\BibitemShut {NoStop}%
\bibitem [{\citenamefont {Renardy}(1996)}]{RenardyYY1996}%
  \BibitemOpen
  \bibfield  {author} {\bibinfo {author} {\bibfnamefont {Y.}~\bibnamefont {Renardy}},\ }\bibfield  {title} {\enquote {\bibinfo {title} {Pattern formation for oscillatory bulk-mode competition in a two-layer b\'enard problem},}\ }\href@noop {} {\bibfield  {journal} {\bibinfo  {journal} {Z angew Math Phys.}\ }\textbf {\bibinfo {volume} {47}},\ \bibinfo {pages} {567--590} (\bibinfo {year} {1996})}\BibitemShut {NoStop}%
\bibitem [{\citenamefont {Degen}, \citenamefont {Colovas},\ and\ \citenamefont {Anderdeck}(1998)}]{Degen1998}%
  \BibitemOpen
  \bibfield  {author} {\bibinfo {author} {\bibfnamefont {M.}~\bibnamefont {Degen}}, \bibinfo {author} {\bibfnamefont {P.}~\bibnamefont {Colovas}}, \ and\ \bibinfo {author} {\bibfnamefont {C.}~\bibnamefont {Anderdeck}},\ }\bibfield  {title} {\enquote {\bibinfo {title} {Time-dependent patterns in the two-layer rayleigh-b\'enard system},}\ }\href@noop {} {\bibfield  {journal} {\bibinfo  {journal} {Phys.\ Rev.\ E}\ }\textbf {\bibinfo {volume} {57(6)}},\ \bibinfo {pages} {5462(11)} (\bibinfo {year} {1998})}\BibitemShut {NoStop}%
\bibitem [{\citenamefont {Neopmnyashchy}\ and\ \citenamefont {Simanovskii}(2004)}]{Neopmnyashchy2004}%
  \BibitemOpen
  \bibfield  {author} {\bibinfo {author} {\bibfnamefont {A.}~\bibnamefont {Neopmnyashchy}}\ and\ \bibinfo {author} {\bibfnamefont {I.}~\bibnamefont {Simanovskii}},\ }\bibfield  {title} {\enquote {\bibinfo {title} {Influence of thermocapillary effect and interfacial heat release on convective oscillations in a two-layer system},}\ }\href@noop {} {\bibfield  {journal} {\bibinfo  {journal} {Phys. Fluids}\ }\textbf {\bibinfo {volume} {16}},\ \bibinfo {pages} {1127--1139} (\bibinfo {year} {2004})}\BibitemShut {NoStop}%
\bibitem [{\citenamefont {Diwakar}\ \emph {et~al.}(2014)\citenamefont {Diwakar}, \citenamefont {Tiwari}, \citenamefont {Das},\ and\ \citenamefont {Sundararajan}}]{Diwakar2014}%
  \BibitemOpen
  \bibfield  {author} {\bibinfo {author} {\bibfnamefont {S.}~\bibnamefont {Diwakar}}, \bibinfo {author} {\bibfnamefont {S.}~\bibnamefont {Tiwari}}, \bibinfo {author} {\bibfnamefont {S.}~\bibnamefont {Das}}, \ and\ \bibinfo {author} {\bibfnamefont {T.}~\bibnamefont {Sundararajan}},\ }\bibfield  {title} {\enquote {\bibinfo {title} {Stability and resonant wave interactions of confined two-layer rayleigh-b\'enard systems},}\ }\href@noop {} {\bibfield  {journal} {\bibinfo  {journal} {J. Fluid Mech.}\ }\textbf {\bibinfo {volume} {754}},\ \bibinfo {pages} {415--455} (\bibinfo {year} {2014})}\BibitemShut {NoStop}%
\bibitem [{\citenamefont {Lyubimova}, \citenamefont {Vorobev},\ and\ \citenamefont {Prokopev}(2019)}]{Lyubimova2019}%
  \BibitemOpen
  \bibfield  {author} {\bibinfo {author} {\bibfnamefont {T.}~\bibnamefont {Lyubimova}}, \bibinfo {author} {\bibfnamefont {A.}~\bibnamefont {Vorobev}}, \ and\ \bibinfo {author} {\bibfnamefont {S.}~\bibnamefont {Prokopev}},\ }\bibfield  {title} {\enquote {\bibinfo {title} {Rayleigh-taylor instability of a miscible interface in a confined domain},}\ }\href@noop {} {\bibfield  {journal} {\bibinfo  {journal} {Phys. Fluids}\ }\textbf {\bibinfo {volume} {31}},\ \bibinfo {pages} {014104} (\bibinfo {year} {2019})}\BibitemShut {NoStop}%
\bibitem [{\citenamefont {Zagvozkin}, \citenamefont {Vorobev},\ and\ \citenamefont {Lyubimova}(2019)}]{Zagvozkin2019}%
  \BibitemOpen
  \bibfield  {author} {\bibinfo {author} {\bibfnamefont {T.}~\bibnamefont {Zagvozkin}}, \bibinfo {author} {\bibfnamefont {A.}~\bibnamefont {Vorobev}}, \ and\ \bibinfo {author} {\bibfnamefont {T.}~\bibnamefont {Lyubimova}},\ }\bibfield  {title} {\enquote {\bibinfo {title} {Kelvin-helmholtz and holmboe instabilities of a diffusive interface between miscible phases},}\ }\href@noop {} {\bibfield  {journal} {\bibinfo  {journal} {Phys. Rev.\ E}\ }\textbf {\bibinfo {volume} {10}},\ \bibinfo {pages} {023103} (\bibinfo {year} {2019})}\BibitemShut {NoStop}%
\bibitem [{\citenamefont {Kheniene}\ and\ \citenamefont {Vorobev}(2015)}]{Kheniene2015}%
  \BibitemOpen
  \bibfield  {author} {\bibinfo {author} {\bibfnamefont {A.}~\bibnamefont {Kheniene}}\ and\ \bibinfo {author} {\bibfnamefont {A.}~\bibnamefont {Vorobev}},\ }\bibfield  {title} {\enquote {\bibinfo {title} {Linear stability of a horizontal phase boundary subjected to shear motion},}\ }\href@noop {} {\bibfield  {journal} {\bibinfo  {journal} {Eur.\ Phys.\ J.\ E}\ }\textbf {\bibinfo {volume} {38}},\ \bibinfo {pages} {77} (\bibinfo {year} {2015})}\BibitemShut {NoStop}%
\bibitem [{\citenamefont {Bestehorn}\ \emph {et~al.}(2021)\citenamefont {Bestehorn}, \citenamefont {Sharma}, \citenamefont {Borcia},\ and\ \citenamefont {Amiroudine}}]{Bestehorn2021}%
  \BibitemOpen
  \bibfield  {author} {\bibinfo {author} {\bibfnamefont {M.}~\bibnamefont {Bestehorn}}, \bibinfo {author} {\bibfnamefont {D.}~\bibnamefont {Sharma}}, \bibinfo {author} {\bibfnamefont {R.}~\bibnamefont {Borcia}}, \ and\ \bibinfo {author} {\bibfnamefont {S.}~\bibnamefont {Amiroudine}},\ }\bibfield  {title} {\enquote {\bibinfo {title} {Faraday instability of binary miscible/immiscible fluids with phase field approach},}\ }\href@noop {} {\bibfield  {journal} {\bibinfo  {journal} {Phys.\ Rev.\ Fluids}\ }\textbf {\bibinfo {volume} {6}},\ \bibinfo {pages} {064002} (\bibinfo {year} {2021})}\BibitemShut {NoStop}%
\bibitem [{\citenamefont {Borcia}\ \emph {et~al.}(2022)\citenamefont {Borcia}, \citenamefont {Borcia}, \citenamefont {Bestehorn}, \citenamefont {Sharma},\ and\ \citenamefont {Amiroudine}}]{Borcia2022}%
  \BibitemOpen
  \bibfield  {author} {\bibinfo {author} {\bibfnamefont {R.}~\bibnamefont {Borcia}}, \bibinfo {author} {\bibfnamefont {I.}~\bibnamefont {Borcia}}, \bibinfo {author} {\bibfnamefont {M.}~\bibnamefont {Bestehorn}}, \bibinfo {author} {\bibfnamefont {D.}~\bibnamefont {Sharma}}, \ and\ \bibinfo {author} {\bibfnamefont {S.}~\bibnamefont {Amiroudine}},\ }\bibfield  {title} {\enquote {\bibinfo {title} {Phase field modeling in liquid binary mixtures: Isothermal and nonisothermal problems},}\ }\href@noop {} {\bibfield  {journal} {\bibinfo  {journal} {Phys.\ Rev.\ Fluids}\ }\textbf {\bibinfo {volume} {7}},\ \bibinfo {pages} {064005} (\bibinfo {year} {2022})}\BibitemShut {NoStop}%
\bibitem [{\citenamefont {Mishra}\ and\ \citenamefont {Diwakar}(2025)}]{Mishra2025}%
  \BibitemOpen
  \bibfield  {author} {\bibinfo {author} {\bibfnamefont {S.}~\bibnamefont {Mishra}}\ and\ \bibinfo {author} {\bibfnamefont {S.}~\bibnamefont {Diwakar}},\ }\bibfield  {title} {\enquote {\bibinfo {title} {Onset of thermo-convective instabilities in two-layer binary fluid systems},}\ }\href@noop {} {\bibfield  {journal} {\bibinfo  {journal} {physics.flu-dyn.}\ }\textbf {\bibinfo {volume} {arXiv:2504.08241}} (\bibinfo {year} {2025})}\BibitemShut {NoStop}%
\bibitem [{\citenamefont {Cahn}\ and\ \citenamefont {Hilliard}(1958)}]{Cahn1958}%
  \BibitemOpen
  \bibfield  {author} {\bibinfo {author} {\bibfnamefont {J.}~\bibnamefont {Cahn}}\ and\ \bibinfo {author} {\bibfnamefont {J.}~\bibnamefont {Hilliard}},\ }\bibfield  {title} {\enquote {\bibinfo {title} {Free energy of a non-uniform system. i. interfacial free energy},}\ }\href@noop {} {\bibfield  {journal} {\bibinfo  {journal} {J. Chem.\ Phys.}\ }\textbf {\bibinfo {volume} {28}},\ \bibinfo {pages} {258} (\bibinfo {year} {1958})}\BibitemShut {NoStop}%
\bibitem [{\citenamefont {Yue}\ \emph {et~al.}(2004)\citenamefont {Yue}, \citenamefont {Feng}, \citenamefont {Liu},\ and\ \citenamefont {Shen}}]{Yue2004}%
  \BibitemOpen
  \bibfield  {author} {\bibinfo {author} {\bibfnamefont {P.}~\bibnamefont {Yue}}, \bibinfo {author} {\bibfnamefont {J.}~\bibnamefont {Feng}}, \bibinfo {author} {\bibfnamefont {C.}~\bibnamefont {Liu}}, \ and\ \bibinfo {author} {\bibfnamefont {J.}~\bibnamefont {Shen}},\ }\bibfield  {title} {\enquote {\bibinfo {title} {A diffuse-interface method for simulating two-phase flows of complex fluids},}\ }\href@noop {} {\bibfield  {journal} {\bibinfo  {journal} {J. Fluid Mech.}\ }\textbf {\bibinfo {volume} {515}},\ \bibinfo {pages} {293--317} (\bibinfo {year} {2004})}\BibitemShut {NoStop}%
\bibitem [{\citenamefont {Ding}, \citenamefont {Spelt},\ and\ \citenamefont {Shu}(2007)}]{Ding2007}%
  \BibitemOpen
  \bibfield  {author} {\bibinfo {author} {\bibfnamefont {H.}~\bibnamefont {Ding}}, \bibinfo {author} {\bibfnamefont {P.}~\bibnamefont {Spelt}}, \ and\ \bibinfo {author} {\bibfnamefont {C.}~\bibnamefont {Shu}},\ }\bibfield  {title} {\enquote {\bibinfo {title} {Diffuse interface model for incompressible two-phase flows with large density ratios},}\ }\href@noop {} {\bibfield  {journal} {\bibinfo  {journal} {J. Comput.\ Phys.}\ }\textbf {\bibinfo {volume} {226}},\ \bibinfo {pages} {2078--2095} (\bibinfo {year} {2007})}\BibitemShut {NoStop}%
\bibitem [{\citenamefont {Ginzburg}\ and\ \citenamefont {Landau}(1950)}]{Ginzburg1950}%
  \BibitemOpen
  \bibfield  {author} {\bibinfo {author} {\bibfnamefont {V.}~\bibnamefont {Ginzburg}}\ and\ \bibinfo {author} {\bibfnamefont {L.}~\bibnamefont {Landau}},\ }\bibfield  {title} {\enquote {\bibinfo {title} {On the thoery of super-conductivity},}\ }\href@noop {} {\bibfield  {journal} {\bibinfo  {journal} {Zh. Eksp. Teor. Fiz.}\ }\textbf {\bibinfo {volume} {20}},\ \bibinfo {pages} {1064} (\bibinfo {year} {1950})}\BibitemShut {NoStop}%
\bibitem [{\citenamefont {Pojman}\ \emph {et~al.}(2006)\citenamefont {Pojman}, \citenamefont {Whitmore}, \citenamefont {Liveri}, \citenamefont {Lombardo}, \citenamefont {Marszalek}, \citenamefont {Parker},\ and\ \citenamefont {Zoltowski}}]{Pojman2006}%
  \BibitemOpen
  \bibfield  {author} {\bibinfo {author} {\bibfnamefont {J.}~\bibnamefont {Pojman}}, \bibinfo {author} {\bibfnamefont {C.}~\bibnamefont {Whitmore}}, \bibinfo {author} {\bibfnamefont {M.}~\bibnamefont {Liveri}}, \bibinfo {author} {\bibfnamefont {R.}~\bibnamefont {Lombardo}}, \bibinfo {author} {\bibfnamefont {J.}~\bibnamefont {Marszalek}}, \bibinfo {author} {\bibfnamefont {R.}~\bibnamefont {Parker}}, \ and\ \bibinfo {author} {\bibfnamefont {B.}~\bibnamefont {Zoltowski}},\ }\bibfield  {title} {\enquote {\bibinfo {title} {Evidence for the existence of an effective interfacial tension between miscible fluids: Isobutyric acid-water and 1-butanol-water in a spinning-drop tensiometer},}\ }\href@noop {} {\bibfield  {journal} {\bibinfo  {journal} {Langmuir}\ }\textbf {\bibinfo {volume} {22}},\ \bibinfo {pages} {2569--2577} (\bibinfo {year} {2006})}\BibitemShut {NoStop}%
\bibitem [{\citenamefont {Jajoo}(2017)}]{Jajoo2017}%
  \BibitemOpen
  \bibfield  {author} {\bibinfo {author} {\bibfnamefont {V.}~\bibnamefont {Jajoo}},\ }\bibfield  {title} {\enquote {\bibinfo {title} {Faraday instability in binary fluids},}\ }\href@noop {} {\bibfield  {journal} {\bibinfo  {journal} {PhD thesis, University of Bordeaux,available online at}\ }\textbf {\bibinfo {volume} {https://tel.archives-ouvertes.fr/tel-01695491}} (\bibinfo {year} {2017})}\BibitemShut {NoStop}%
\bibitem [{\citenamefont {Jacqmin}(1999)}]{Jacqmin1999}%
  \BibitemOpen
  \bibfield  {author} {\bibinfo {author} {\bibfnamefont {D.}~\bibnamefont {Jacqmin}},\ }\bibfield  {title} {\enquote {\bibinfo {title} {Calculation of two-phase navier-stokes flows using phase-field modeling},}\ }\href@noop {} {\bibfield  {journal} {\bibinfo  {journal} {J. Comput.\ Phys.}\ }\textbf {\bibinfo {volume} {155}},\ \bibinfo {pages} {96--127} (\bibinfo {year} {1999})}\BibitemShut {NoStop}%
\bibitem [{\citenamefont {Diwakar}\ \emph {et~al.}(2018)\citenamefont {Diwakar}, \citenamefont {Jajoo}, \citenamefont {Amiroudine}, \citenamefont {Matsumoto}, \citenamefont {Narayanan},\ and\ \citenamefont {Zoueshtiagh}}]{Diwakar2018}%
  \BibitemOpen
  \bibfield  {author} {\bibinfo {author} {\bibfnamefont {S.}~\bibnamefont {Diwakar}}, \bibinfo {author} {\bibfnamefont {V.}~\bibnamefont {Jajoo}}, \bibinfo {author} {\bibfnamefont {S.}~\bibnamefont {Amiroudine}}, \bibinfo {author} {\bibfnamefont {S.}~\bibnamefont {Matsumoto}}, \bibinfo {author} {\bibfnamefont {R.}~\bibnamefont {Narayanan}}, \ and\ \bibinfo {author} {\bibfnamefont {F.}~\bibnamefont {Zoueshtiagh}},\ }\bibfield  {title} {\enquote {\bibinfo {title} {Influence of capillarity and gravity on confined faraday waves},}\ }\href@noop {} {\bibfield  {journal} {\bibinfo  {journal} {Phys.\ Rev.\ Fluids}\ }\textbf {\bibinfo {volume} {3}},\ \bibinfo {pages} {073902} (\bibinfo {year} {2018})}\BibitemShut {NoStop}%
\bibitem [{\citenamefont {Abels}, \citenamefont {Depner},\ and\ \citenamefont {Garcke}(2013)}]{Abels2013}%
  \BibitemOpen
  \bibfield  {author} {\bibinfo {author} {\bibfnamefont {H.}~\bibnamefont {Abels}}, \bibinfo {author} {\bibfnamefont {D.}~\bibnamefont {Depner}}, \ and\ \bibinfo {author} {\bibfnamefont {H.}~\bibnamefont {Garcke}},\ }\bibfield  {title} {\enquote {\bibinfo {title} {On an incompressible navier-stokes/cahn-hilliard system with degenrate mobility},}\ }\href@noop {} {\bibfield  {journal} {\bibinfo  {journal} {Ann. Inst. Henri Poincare}\ }\textbf {\bibinfo {volume} {30}},\ \bibinfo {pages} {1175--1190} (\bibinfo {year} {2013})}\BibitemShut {NoStop}%
\bibitem [{\citenamefont {Joseph}, \citenamefont {Huang},\ and\ \citenamefont {Hu}(1996)}]{Joseph1996}%
  \BibitemOpen
  \bibfield  {author} {\bibinfo {author} {\bibfnamefont {D.}~\bibnamefont {Joseph}}, \bibinfo {author} {\bibfnamefont {A.}~\bibnamefont {Huang}}, \ and\ \bibinfo {author} {\bibfnamefont {H.}~\bibnamefont {Hu}},\ }\bibfield  {title} {\enquote {\bibinfo {title} {Non-solenoidal velocity effects and korteweg stresses in simple mixtures of incompressible liquids},}\ }\href@noop {} {\bibfield  {journal} {\bibinfo  {journal} {Physica D}\ }\textbf {\bibinfo {volume} {97}},\ \bibinfo {pages} {104--125} (\bibinfo {year} {1996})}\BibitemShut {NoStop}%
\bibitem [{\citenamefont {Magaletti}\ \emph {et~al.}(2013)\citenamefont {Magaletti}, \citenamefont {Picano}, \citenamefont {Chinappi}, \citenamefont {Marino},\ and\ \citenamefont {Casciola}}]{Magaletti2013}%
  \BibitemOpen
  \bibfield  {author} {\bibinfo {author} {\bibfnamefont {F.}~\bibnamefont {Magaletti}}, \bibinfo {author} {\bibfnamefont {F.}~\bibnamefont {Picano}}, \bibinfo {author} {\bibfnamefont {M.}~\bibnamefont {Chinappi}}, \bibinfo {author} {\bibfnamefont {L.}~\bibnamefont {Marino}}, \ and\ \bibinfo {author} {\bibfnamefont {C.}~\bibnamefont {Casciola}},\ }\bibfield  {title} {\enquote {\bibinfo {title} {The sharp-interface limit of the cahn-hilliard/navier-stokes model for binary fluids},}\ }\href@noop {} {\bibfield  {journal} {\bibinfo  {journal} {J. Fluid Mech.}\ }\textbf {\bibinfo {volume} {714}},\ \bibinfo {pages} {95--126} (\bibinfo {year} {2013})}\BibitemShut {NoStop}%
\bibitem [{\citenamefont {Tee}\ and\ \citenamefont {Trefethen}(2006)}]{Tee2006}%
  \BibitemOpen
  \bibfield  {author} {\bibinfo {author} {\bibfnamefont {T.}~\bibnamefont {Tee}}\ and\ \bibinfo {author} {\bibfnamefont {L.}~\bibnamefont {Trefethen}},\ }\bibfield  {title} {\enquote {\bibinfo {title} {A rational spectral collocation method with adaptively transformed chebyshev grid points},}\ }\href@noop {} {\bibfield  {journal} {\bibinfo  {journal} {Siam.\ J.\ Sci.\ Comput}\ }\textbf {\bibinfo {volume} {28}},\ \bibinfo {pages} {1798--1811} (\bibinfo {year} {2006})}\BibitemShut {NoStop}%
\bibitem [{\citenamefont {Diwakar}\ \emph {et~al.}(2015)\citenamefont {Diwakar}, \citenamefont {Zoueshtiagh}, \citenamefont {Amiroudine},\ and\ \citenamefont {Narayanan}}]{Diwakar2015}%
  \BibitemOpen
  \bibfield  {author} {\bibinfo {author} {\bibfnamefont {S.}~\bibnamefont {Diwakar}}, \bibinfo {author} {\bibfnamefont {F.}~\bibnamefont {Zoueshtiagh}}, \bibinfo {author} {\bibfnamefont {S.}~\bibnamefont {Amiroudine}}, \ and\ \bibinfo {author} {\bibfnamefont {R.}~\bibnamefont {Narayanan}},\ }\bibfield  {title} {\enquote {\bibinfo {title} {The faraday instability in miscible fluid systems},}\ }\href@noop {} {\bibfield  {journal} {\bibinfo  {journal} {Phys. Fluids}\ }\textbf {\bibinfo {volume} {27}},\ \bibinfo {pages} {084111} (\bibinfo {year} {2015})}\BibitemShut {NoStop}%
\bibitem [{\citenamefont {Renardy}\ and\ \citenamefont {Stoltz}(2000)}]{Renardy2000}%
  \BibitemOpen
  \bibfield  {author} {\bibinfo {author} {\bibfnamefont {Y.}~\bibnamefont {Renardy}}\ and\ \bibinfo {author} {\bibfnamefont {C.}~\bibnamefont {Stoltz}},\ }\bibfield  {title} {\enquote {\bibinfo {title} {Time-dependent pattern formation for convection in two layers of immiscible liquids},}\ }\href@noop {} {\bibfield  {journal} {\bibinfo  {journal} {Int.\ J.\ Multiph.\ Flow}\ }\textbf {\bibinfo {volume} {26}},\ \bibinfo {pages} {1875--1889} (\bibinfo {year} {2000})}\BibitemShut {NoStop}%
\end{thebibliography}%

\end{document}